\documentclass[12pt]{article}
\usepackage{amssymb,amsmath}
\usepackage{color}
\numberwithin{equation}{section}
\newtheorem{theorem}{Theorem}[section]     

\newtheorem{defi}[theorem]{Definition}
\newtheorem{proposition}[theorem]{Proposition}
\newtheorem{lemma}[theorem]{Lemma}

\newtheorem{remark}[theorem]{Remark}

\def\d{\partial}
\def\n{\noindent}
\def\f{\frac}

\def\proof{\noindent\hspace{2em}{\itshape Proof: }}
\def\QEDclosed{\mbox{\rule[0pt]{1.3ex}{1.3ex}}} 

\def\QED{\QEDclosed} 
\def\endproof{\hspace*{\fill}~\QED\par\endtrivlist\unskip}
\newcommand{\eqa}{\begin{eqnarray}}
\newcommand{\eeqa}{\end{eqnarray}}
\newcommand{\beq}{\begin{equation}}
\newcommand{\eeq}{\end{equation}}

\begin{document}
\title{From Darboux-Egorov system\\ to bi-flat $F$-manifolds}
\author{Alessandro Arsie* and Paolo Lorenzoni**\\
\\
{\small *Department of Mathematics and Statistics}\\
{\small University of Toledo,}
{\small 2801 W. Bancroft St., 43606 Toledo, OH, USA}\\
{\small **Dipartimento di Matematica e Applicazioni}\\
{\small Universit\`a di Milano-Bicocca,}
{\small Via Roberto Cozzi 53, I-20125 Milano, Italy}\\
{\small *alessandro.arsie@utoledo.edu,  **paolo.lorenzoni@unimib.it}}

\date{}

\maketitle

\begin{abstract}
Motivated by the theory of integrable PDEs of hydrodynamic type and by the generalization of Dubrovin's duality
 in the framework of $F$-manifolds due to Manin \cite{manin}, we consider a special class of $F$-manifolds, called bi-flat $F$-manifolds.
 
 A bi-flat $F$-manifold is given by the following data $(M, \nabla_1,\nabla_2,\circ,*,e,E)$,
 where $(M, \circ)$ is an $F$-manifold, $e$ is the identity of the product $\circ$, $\nabla_1$
 is a flat connection compatible with $\circ$ and satisfying $\nabla_1 e=0$, while $E$ is an eventual identity giving rise to the dual product
 $*$, and $\nabla_2$ is a flat connection compatible with $*$ and satisfying $\nabla_2 E=0$.
 Moreover, the two connections $\nabla_1$ and $\nabla_2$ are required to be hydrodynamically
 almost equivalent in the sense specified in \cite{AL2012}. 

First we show that, similarly to the way in which Frobenius manifolds are constructed starting from Darboux-Egorov systems, also bi-flat $F$-manifolds can be built from solutions of suitably augmented Darboux-Egorov systems, essentially dropping the requirement that the rotation coefficients are symmetric. 

Although any Frobenius manifold possesses automatically the structure of a bi-flat $F$-manifold, we show that the latter is a strictly larger class.

In particular we study in some detail bi-flat $F$-manifolds in dimensions $n=2, 3$. For instance, we show that in dimension three bi-flat $F$-manifolds are parametrized by solutions of a two parameters Painlev\'e VI equation, admitting among its solutions hypergeometric functions. 
Finally we comment on some open problems of wide scope related to bi-flat $F$-manifolds.

\end{abstract}
\section{Introduction}

Since its resurgence at the end of the sixties, the theory of integrable systems, expanded to include infinite dimensional systems has provided different perspectives and new powerful tools to several areas in mathematics, both pure and applied. To name a few, just consider the development of the Inverse Scattering Transform and its applications or the celebrated Witten's conjecture, proved by Kontsevich, which states that the generating function for intersection numbers of Mumford-Morita-Miller stable classes on the Deligne-Mumford compactification of moduli space of pointed curves is related to a suitable tau function for the KdV integrable hierarchy. 

Another milestone in this development was the introduction by Dubrovin, at the beginning of the nineties, of the so called Frobenius manifolds, mainly as a way to provide an intrinsic geometric description of the properties of WDDV equations and to formalize the mathematical structure of the genus zero Gromov-Witten invariants of complex projective (or compact symplectic) manifolds (see \cite{du93}). It turned out that Frobenius manifolds are a convenient framework capable of unifying apparently very different theories, ranging from singularity theory (unfolding spaces of isolated hypersurface singularities), to important classes of integrable PDEs (integrable PDEs of hydrodynamic type) and phenomena in geometry (for instance the Barannikov-Kontsevich construction of Frobenius manifolds starting with the Dolbeault complex of a Calabi-Yau manifold).

In the last twenty years Frobenius manifolds have been the subject of several investigations, often leading to an expansion of the conceptual framework in which they were born. In this direction, Hertling and Manin (see \cite{HM}) introduced weak Frobenius manifolds, i.e. Frobenius manifolds without a pre-fixed flat metric to relate the standard construction of Frobenius manifolds to those provided by K. Sato and Barannikov-Kontsevich. In the same paper, Hertling and Manin introduced also the notion of $F$-manifold, whose definition we are going to recall below. 

An \emph{$F$-manifold} is a manifold $M$ endowed with an associative commutative product $\circ$ on vector fields $X$, $Y$:
$$
(X\circ Y)^i:=c^i_{jk}X^j Y^k\ ,
$$
satisfying the condition
\beq
\label{hmcomponents}
(\d_s c^k_{jl})c^s_{im}+(\d_j c^s_{im})c^k_{sl}-(\d_s c^k_{im})c^s_{jl}-(\d_i c^s_{jl})c^k_{sm}-(\d_l c^s_{jm})c^k_{si}-(\d_m c^s_{li})c^k_{js}=0\ .
\eeq
The above condition is known as \emph{Hertling-Manin condition}. One usually requires
 also the existence of a special vector field $e$, called \emph{unit vector field} satisfying the
 condition
$$X\circ e=X$$
for any vector field $X$. 
 If there exist special coordinates, called \emph{canonical coordinates}, such that  $$c^i_{jk}=\delta^i_j\delta^i_k$$
the $F$-manifold is said to be \emph{semisimple}.

Since $F$-manifolds in general are not equipped with a metric, it is in principle difficult to relate
 them to systems of PDEs of hydrodynamic type, especially after it has become more and more apparent how the study of
 these systems leads naturally to some classical problems in Riemannian geometry,
 starting from the pioneering work of Dubrovin and Novikov \cite{dn84}. 

Pursuing a point of view initiated in \cite{LP} and \cite{LPR} and further expanded in \cite{AL2012}, in this paper we will focus our attention on a class of $F$-manifolds equipped with a pair of compatible flat connections not necessarily originating from a metric, to which many of the construction available for Frobenius manifolds can be indeed generalized.
 
 First of all, following Manin,
 we recall the definition of $F$-manifolds with compatible flat structure:
\begin{defi}\cite{manin}
A (semisimple) $F$-manifold $(M,\nabla,\circ,e)$ with compatible flat structure
 (shortly a flat $F$-manifold)
 is a (semisimple) $F$-manifold $(M,\circ,e)$ endowed with a flat connections $\nabla$ compatible with the product $\circ$, i.e.
\beq\label{scc}
\nabla_l c^i_{jk}=\nabla_j c^i_{lk}
\eeq
and satisfying the condition $\nabla e=0$.
\end{defi}  

Secondly, we recall the definition of eventual identities \cite{manin} (see also \cite{DS})
 that generalizes the almost dual structure of Frobenius manifolds \cite{Dad}.

\begin{defi}
A vector field E on an $F$-manifold is called an eventual identity, if it is invertible with respect to $\circ$
 (i.e. there is a vector field $E^{-1}$ such that
$E\circ E^{-1} = E^{-1} \circ E = e$) and, moreover, the bilinear product $*$ defined via
\beq\label{nm}
X *Y := X \circ Y \circ E^{-1},\qquad \text{ for all } X, Y \text{ vector fields}
\eeq
defines a new $F$-manifold structure on M.
\end{defi}

The presence of a second product leads naturally to consider the concept of duality in the framework of
 $F$-manifolds with compatible flat structure \cite{manin} (see also \cite{DS2}). Taking into account
 the notion of hydrodynamically equivalent connections introduced in \cite{AL2012} we have the following 
 definition.

\begin{defi}\label{defibiflat}
A \emph{bi-flat} (semisimple) $F$-manifold $(M,\nabla_1,\nabla_2,\circ,*,e,E)$
 is a (semisimple) $F$-manifold $(M,\circ,e)$ endowed with a pair
 of flat connections $\nabla_1$ and $\nabla_2$ and with an eventual identity $E$ satisfying the following conditions:
\begin{itemize}
\item $\nabla_1$ is compatible with the product $\circ$ 
and $\nabla_1 e=0$,
\item $\nabla_2$ is compatible with the product $*$ and $\nabla_2 E=0$,
\item $\nabla_1$ and $\nabla_2$ are almost hydrodynamically equivalent  i.e.
\beq\label{almostcomp}
(d_{\nabla_1}-d_{\nabla_2})(X\,\circ)=0,\qquad{\rm or}\qquad(d_{\nabla_1}-d_{\nabla_2})(X\,*)=0
\eeq
for every vector fields $X$; here $d_{\nabla}$ is the exterior covariant derivative \footnote{See for instance \cite{LEE}, page 536.}
 constructed from a connection $\nabla$.  
\end{itemize}
\end{defi}
Let us remark that equations \eqref{almostcomp} simply mean that in canonical coordinates for $\circ$ one has
 $\stackrel{1}{\Gamma^i}_{ij}-\stackrel{2}{\Gamma^i}_{ij}=0$, $i\neq j$, where $\stackrel{1}{\Gamma}$ and $\stackrel{2}{\Gamma}$
 are the Christoffel symbols associated to the connections $\nabla_1$ and $\nabla_2$ respectively. An analogous statement holds
 true in canonical coordinates for $*$,  see \cite{AL2012}.

Observe that any Frobenius manifold possesses automatically a bi-flat $F$-manifold structure in the above sense. The second flat connection $\nabla_2$
 is hydrodymically equivalent \footnote{The difference between the Christoffel synmbols
 is proportional to the structure constants.} to
 the Levi-Civita connection $\nabla$ of the intersection form
 which, as it is well-known, does not fulfill the condition $\nabla E=0$.
 However, as it will be clear through the examples explored in this paper (see Section 5, 6 and 7), bi-flat $F$-manifolds
 are a genuine generalization of Frobenius manifolds. Bi-flat $F$-manifolds therefore lie between Frobenius manifolds
 and $F$-manifolds in terms of generality, but as we are going to see they appear to be particularly well suited to deal
 with integrable PDEs of hydrodynamic type. 

Indeed, as shown in \cite{LPR} any $F$-manifold with compatible flat structure defines an integrable
 hierarchy of hydrodynamic type. The presence of a second flat structure is related to a recursive scheme to
 produce the flows of the hierarchy \cite{AL2012} providing a nontrivial generalization of the usual bi-Hamiltonian scheme. In this sense, bi-flat $F$-manifolds appear a convenient framework to deal with integrable hierarchies of hydrodynamic type,
 encompassing also examples, like the $\epsilon$-system that are beyond the usual Frobenius set-up. 
We comment briefly on the relationships between bi-flat $F$-manifolds and integrable hierarchies in Section 8.
 
One of the aims of our paper is to show that $F$-manifolds with compatible
 bi-flat structure can be constructed starting from solutions of a differential
 system of Darboux-Egorov type in a way similar to what it is done for Frobenius manifolds (see \cite{du93}, Lecture 3),
 substantially dropping the requirement that the rotation coefficients $\beta_{ij}$ are symmetric. 
  
Indeed, we are going to show how to construct $F$-manifolds with compatible bi-flat structure using the
solutions of the system:
\begin{eqnarray}
\label{ED1}
&&\d_k\beta_{ij}=\beta_{ik}\beta_{kj},\qquad k\ne i\ne j\ne k\\
\label{ED2}
&&e(\beta_{ij})=0,\\
\label{ED3}
&&E(\beta_{ij})=-\beta_{ij},
\end{eqnarray}
where $e=\sum\f{\d}{\d u^i}$, $E=\sum_i u^i\f{\d}{\d u^i}$. The functions $\beta_{ij}$ are usually known as Ricci's rotation coefficients (see \cite{Darboux} and \cite{Ricci}) in literature. 

In the symmetric case $\beta_{ij}=\beta_{ji}$ such
 relations have been studied in details by Dubrovin in \cite{du93,du99}.
 In this case the solutions of the system (\ref{ED1},\ref{ED2},\ref{ED3}) are related to Frobenius manifolds and
  the system (\ref{ED1},\ref{ED2}) is called \emph{Darboux-Egorov system} (see \cite{Darboux, Egorov}).\footnote{Darboux in his treatise \cite{Darboux} already considered system of the form \ref{ED1} with $\beta_{ij}$ not symmetric but he augmented the system with other conditions that assure that the $\beta_{ij}$ originate from a flat curvilinear coordinates system.}

If $\beta_{ij}$ are not symmetric,
 starting from the solutions
 $\beta_{ij}(u^1,\dots,u^n)$  of the system (\ref{ED1},\ref{ED2},\ref{ED3}) it is possible to construct flat and bi-flat
 semisimple $F$-manifolds.  
Let us also underline that it is by no means obvious how to construct such flat structures. Indeed, although starting from $\beta_{ij}$
 one can still introduce Lam\'e coefficients and the corresponding diagonal metrics, the flat connections we are looking for are not
 the Levi-Civita connections of these metrics. Instead, the flat connections are obtained imposing their compatibility with respect
 to two different product structures. 

In our opinion it is very surprising that in the intensively studied classical system (\ref{ED1},\ref{ED2},\ref{ED3})
 associated to flat curvilinear coordinate systems there is hidden a deep relationship to the theory of $F$-manifolds,
 based on the compatibility with product structures. 


 Furthermore, we give a straightforward and complete proof of the fact that in the case $n=3$ the system (\ref{ED1},\ref{ED2},\ref{ED3})
 is equivalent to a two-parameter Painlev\'e VI equation, thus proving that three-dimensional bi-flat $F$-manifolds are parametrized by solutions of a two parameters Painlev\'e VI equation.

\vspace{0.2in}
The paper is organized as follows. In Section 2, we show that, although the solutions of the system  (\ref{ED1},\ref{ED2}) no longer define flat diagonal metrics in the case in which $\beta_{ij}\neq \beta_{ji}$, it is still possible to define a flat connection $\nabla_1$ that satisfies $\nabla_1 e=0$. We call $\nabla_1$ the natural connection. 

In Section 3 we show that the system (\ref{ED1},\ref{ED2}) augmented with \eqref{ED3} is a compatible system
and we provide a construction for a second flat connection $\nabla_2$ satisfying $\nabla_2E=0$. In Section 4, combining the results obtained in previous sections we show how to construct bi-flat semisimple $F$-manifolds $(\nabla_1, \nabla_2, \circ, \star, e, E)$ starting from solutions of (\ref{ED1}, \ref{ED2}, \ref{ED3}). 


In Section 5 we construct concrete examples of bi-flat $F$-manifolds in dimension $n=2$. In Section 6 we show that bi-flat $F$-manifolds in dimension $n=3$ are parametrized by solutions of a two-parameter  Painlev\'e VI equation; this should be contrasted to the case of Frobenius manifolds, that for $n=3$ are described by the solutions of a single parameter Painlev\'e VI equation that never admits hypergeometric functions as solutions. In Section 7 we consider a class of examples of bi-flat $F$-manifolds in arbitrary dimensions $n>2$, which are constructed starting from the  $\epsilon$-system.
 
In the final Section 8, we provide some conclusions and perspectives on open problems of wide scope.
\section{The natural connections}
In  the symmetric case $\beta_{ij}=\beta_{ji}$, 
the solutions $\beta_{ij}(u^1,\dots,u^n)$ ($i\ne j$)  of the system \eqref{ED1} and \eqref{ED2}
 (usually called  \emph{rotation coefficients}) describe diagonal flat metrics. More precisely, given some rotation coefficients satisfying \eqref{ED1} and \eqref{ED2}, any
 solution $(H_1,\dots,H_n)$ (the so-called Lam\'{e} coefficients) of the system 
\begin{eqnarray}
\label{L1}
&&\d_j H_i=\beta_{ij}H_j,\qquad i\ne j
\end{eqnarray}
defines a flat diagonal metric $g_{ii}=H_i^2$. In particular one can choose among the solutions of \eqref{L1} those satisfying the further condition
\begin{eqnarray}
\label{L2}
&&e(H_i)=0,
\end{eqnarray}
In the non symmetric case  the solutions of the system (\ref{ED1},\ref{ED2}) no longer define flat diagonal metrics. However is still possible to define a flat connection.

\begin{remark}
Both the systems (\ref{ED1},\ref{ED2}) and  (\ref{L1},\ref{L2}) (given $\beta_{ij}$ satisfying  (\ref{ED1},\ref{ED2}))
 are compatible. The proof is a straightforward (not short) computation.
\end{remark}

\begin{theorem}\label{naturalnabla}
The symmetric  connection  $\nabla_1$ defined by
\begin{equation}\label{naturalc}
\begin{split}
\Gamma^i_{jk}&:=0\qquad\forall i\ne j\ne k \ne i\\
\Gamma^i_{jj}&:=-\Gamma^i_{ij}\qquad i\ne j\\
\Gamma^i_{ij}&:=\f{H_j}{H_i}\beta_{ij}\qquad i\ne j\\
\Gamma^i_{ii}&:=-\sum_{l\ne i}\Gamma^i_{li}
\end{split}
\end{equation}
where $\beta_{ij}$ ($i\ne j$) and $H_i$ are solutions of the systems (\ref{ED1},\ref{ED2}) and (\ref{L1},\ref{L1}), is flat.
\end{theorem}

\n
{\emph Proof}. For distinct indices clearly the components of the Riemann tensor vanish. For $i\ne j\ne k\ne i$:
\begin{eqnarray*}
&&R^i_{jki}=-R^i_{jik}=\d_k\Gamma^i_{ij}+\Gamma^i_{ik}\Gamma^i_{ij}-\Gamma^i_{ij}\Gamma^j_{jk}
-\Gamma^i_{ik}\Gamma^k_{kj}=\\
&&\d_k\left(\f{H_j}{H_i}\beta_{ij}\right)+\f{H_k H_j}{(H_i)^2}\beta_{ik}\beta_{ij}-\f{H_k}{H_i}\beta_{ij}\beta_{jk}
-\f{H_j}{H_i}\beta_{ik}\beta_{kj}=\\
&&\f{\d_k H_j}{H_i}\beta_{ij}-\f{H_j\d_k H_i}{(H_i)^2}\beta_{ij}+\f{H_j}{H_i}\beta_{ik}\beta_{kj}+\f{H_k H_j}{(H_i)^2}\beta_{ik}\beta_{ij}-\f{H_k}{H_i}\beta_{ij}\beta_{jk}
-\f{H_j}{H_i}\beta_{ik}\beta_{kj}=0.
\end{eqnarray*}
The vanishing of
\begin{equation*}
R^i_{jjk}=-R^i_{jkj}=-\d_k\Gamma^i_{jj}
+\Gamma^i_{jj}\Gamma^j_{jk}
-\Gamma^i_{ik}\Gamma^i_{jj}
-\Gamma^i_{kk}\Gamma^k_{jj}=0\,\,\,\mbox{if $i\ne k\ne j\ne i$}
\end{equation*}
follows from the vanishing of $R^i_{jki}$ and from 
 $\Gamma^i_{jj}=-\Gamma^i_{ij}$. Indeed using the
 last identity we obtain
\begin{equation*}
R^i_{jjk}=\d_k\Gamma^i_{ij}
-\Gamma^i_{ij}\Gamma^j_{jk}
+\Gamma^i_{ik}\Gamma^i_{ij}
-\Gamma^i_{ik}\Gamma^k_{kj}=R^i_{jki}=0.
\end{equation*}  
The vanishing of  
$$R^i_{ikl}=-R^i_{ilk}=\d_k\Gamma^i_{il}-\d_l\Gamma^i_{ik}=0\,\,\,\mbox{if $i\ne k\ne l\ne i$}$$
is immediate using the vanishing of $R^i_{jki}$ and the contracted first Bianchi identity $R^i_{ijk}+R^i_{jki}+R^i_{kij}=0.$

Finally, for $i\ne j$ we have
\begin{eqnarray*}
&&R^i_{iji}=-R^i_{iij}=\d_j\Gamma^i_{ii}-\d_i\Gamma^i_{ij}=
-\d_j\sum_{l\ne i}\left(\f{H_l}{H_i}\beta_{il}\right)
-\d_i\left(\f{H_j}{H_i}\beta_{ij}\right)=\\
&&-\d_j\left(\f{H_j}{H_i}\beta_{ij}\right)
-\d_j\sum_{l\ne i,j}\left(\f{H_l}{H_i}\beta_{il}\right)
-\d_i\left(\f{H_j}{H_i}\beta_{ij}\right)=\\
&&-\f{\d_j H_j}{H_i}\beta_{ij}
+H_j\f{\d_j H_i}{(H_i)^2}\beta_{ij}
-\f{H_j}{H_i}\d_j \beta_{ij}
-\sum_{l\ne i,j}\left(\f{\d_j H_l}{H_i}\beta_{il}\right)
+\sum_{l\ne i,j}\left(H_l\f{\d_j H_i}{(H_i)^2}\beta_{il}\right)+\\
&&-\sum_{l\ne i,j}\left(\f{H_l}{H_i}\d_j\beta_{il}\right)
-\f{\d_i H_j}{H_i}\beta_{ij}
+H_j\f{\d_i H_i}{(H_i)^2}\beta_{ij}
-\f{H_j}{H_i}\d_i\beta_{ij}=\\
&&\sum_{l\ne i,j}\f{H_l}{H_i}\beta_{jl}\beta_{ij}
+\f{(H_j)^2}{(H_i)^2}(\beta_{ij})^2
-\f{H_j}{H_i}(\d_j+\d_i)\beta_{ij}
-\sum_{l\ne i,j}\left(\f{H_j}{H_i}\beta_{lj}\beta_{il}\right)
+\sum_{l\ne i,j}\left(\f{H_j H_l}{(H_i)^2}\beta_{ij}\beta_{il}\right)+\\
&&-\sum_{l\ne i,j}\left(\f{H_l}{H_i}\d_j\beta_{il}\right)
-\f{H_j}{(H_i)^2}\beta_{ij}\sum_{l\ne i}\d_l H_i=\\
&&\sum_{l\ne i}\left(\f{H_j H_l}{(H_i)^2}\beta_{ij}\beta_{il}\right)-\f{H_j}{(H_i)^2}\beta_{ij}\sum_{l\ne i}\d_l H_i=0,
\end{eqnarray*}
and
\begin{eqnarray*}
&&R^i_{jji}=-R^i_{jij}=\d_j\Gamma^i_{ji}-\d_i\Gamma^i_{jj}
+\left(\Gamma^i_{ij}\right)^2
+\Gamma^i_{jj}\Gamma^j_{ji}
-\sum_{l=1}^n\Gamma^i_{li}\Gamma^l_{jj}=\\
&&(\d_j+\d_i)\Gamma^i_{ji}+
\Gamma^i_{ij}\left(\Gamma^i_{ij}-\Gamma^j_{ji}
+\Gamma^i_{ii}-\Gamma^j_{jj}\right)+\sum_{l\ne i,j}^n\Gamma^i_{li}\Gamma^l_{lj}=\\
&&(\d_j+\d_i)\Gamma^i_{ji}+
\Gamma^i_{ij}\left(-\sum_{l\ne i,j}\Gamma^i_{il}+\sum_{l\ne i,j}\Gamma^j_{lj}\right)
+\sum_{l\neq i,j}^n\Gamma^i_{li}\Gamma^l_{lj}=\\
&&\beta_{ij}(\d_i+\d_j)\f{H_j}{H_i}-\f{H_j}{H_i}\sum_{l\neq i,j}\d_l\beta_{ij}+\f{H_j}{H_i}\beta_{ij}\left(-\sum_{l\ne i,j}\f{H_l}{H_i}\beta_{il}+\sum_{l\ne i,j}\f{H_l}{H_j}\beta_{jl}\right)
+\sum_{l\ne i,j}^n\f{H_j}{H_i}\beta_{il}\beta_{lj}=\\
&&\beta_{ij}\left(\f{\d_i H_j}{H_i}+\f{\d_j H_j}{H_i}
-\f{H_j\d_i H_i}{(H_i)^2}-\f{H_j\d_j H_i}{(H_i)^2}\right)
+\f{H_j}{H_i}\beta_{ij}\left(-\sum_{l\ne i,j}\f{H_l}{H_i}\beta_{il}+\sum_{l\ne i,j}\f{H_l}{H_j}\beta_{jl}\right)=\\
&&\beta_{ij}\left(-\f{\sum_{l\ne i,j}\d_l H_j}{H_i}
+\f{H_j\sum_{l\ne i,j}\d_l H_i}{(H_i)^2}\right)
+\f{H_j}{H_i}\beta_{ij}\left(-\sum_{l\ne i,j}\f{H_l}{H_i}\beta_{il}+\sum_{l\ne i,j}\f{H_l}{H_j}\beta_{jl}\right)=0.\\
\end{eqnarray*}
This proves the claim.
\endproof

\begin{lemma}\label{lemmac}
The connection $\nabla_1$ defined in \eqref{naturalc} is compatible with the product $c^i_{jk}=\delta^i_j\delta^i_k$ and satisfies $\nabla_1 e=0$. 
\end{lemma}
\proof
The fact that $\nabla_1 e=0$ identically is equivalent to the requirements $\Gamma^i_{jj}:=-\Gamma^i_{ij}$
 for $i\ne j$ and $\Gamma^i_{ii}:=-\sum_{l\ne i}\Gamma^i_{li}$ that appear in \eqref{naturalc}. 

Moreover, the compatibility of $\nabla_1$  with $c^i_{jk}:=\delta^i_j\delta^i_k$ is equivalent
 to the requirements  $\Gamma^i_{jj}:=-\Gamma^i_{ij}$ for $i\ne j$
 and $\Gamma^i_{jk}=0$ for $i\ne j\ne k\ne i$. All these statements can be checked
 via straightforward computations in coordinates $\{u^1, \dots, u^n\}$. The Lemma is proved.
\endproof



\begin{defi}
Following \cite{LP} we call $\nabla_1$ \emph{the natural connection} associated with $\beta_{ij}$ and $H_i$. 
\end{defi}

Summarizing, combining Theorem \ref{naturalnabla} and Lemma \ref{lemmac} we have proved the following:
\begin{theorem}\label{mth1}
Let $(\beta_{ij},H_i)$ be a solution of the system (\ref{ED1},\ref{ED2}) and (\ref{L1}\ref{L2}), then 
 the natural connection $\nabla_1$ associated with $(\beta_{ij},H_i)$ and the structure constants  defined in the 
 coordinates $(u^1,\dots,u^n)$ by $c^i_{jk}:=\delta^i_j\delta^i_k$ give rise to an $F$-manifold with compatible  flat structure. 
\end{theorem}

\section{The dual connections}

\begin{proposition}
The system (\ref{ED1},\ref{ED2}) augmented with the equation \eqref{ED3} is a complete compatible system. 
\end{proposition}

\proof
The completeness follows because each derivative of the functions $\beta_{ij}$ is specified with respect to all variables. Indeed, using the equations $E(\beta_{ij})=-\beta_{ij}$ and $e(\beta_{ij})=0$ it is easy to get the following equations:
$$\partial_i\beta_{ij}=\frac{1}{u^j-u^i}\left\{ \sum_{l\neq i,j}(u^l-u^j)\partial_l \beta_{ij}+\beta_{ij} \right\},$$
$$\partial_j\beta_{ij}=\frac{1}{u^j-u^i}\left\{\sum_{l\neq i,j}(u^i-u^l)\partial_l \beta_{ij}-\beta_{ij} \right\}.$$
The compatibility follows checking that $\partial_i \partial_k\beta_{ij}-\partial_k\partial_i \beta_{ij}=0$ identically,  and checking that 
$\partial_i \partial_j \beta_{ij}-\partial_j \partial_i \beta_{ij}=0$ identically. These are straightforward long computations. 
\endproof
By the previous Proposition it follows that the general solution of the system given by \eqref{ED1}, \eqref{ED2} and \eqref{ED3} depends on $n(n-1)$ arbitrary constants. 

In this section we show that, starting from a solution of  (\ref{ED1},\ref{ED3}), one can costruct a second flat 
 connection. Before illustrating the costruction let us observe that, due to (\ref{ED1},\ref{ED3}) the equation
\begin{equation}
\label{L3}
E(H_i)=-dH_i
\end{equation}
is compatible with \eqref{L1}. Taking into account this fact we can prove the following.

\begin{theorem}\label{dualnablaflathm}
If $\beta_{ij}$ ($i\ne j$) and $H_i$ are solutions of the systems (\ref{ED1},\ref{ED3}) and (\ref{L1},\ref{L3}) respectively, the connection $\nabla_2$ defined by
\begin{equation}\label{dualnabla}
\begin{split}
\Gamma^i_{jk}&:=0\qquad\forall i\ne j\ne k \ne i\\
\Gamma^i_{jj}&:=-\f{u^i}{u^j}\Gamma^i_{ij}\qquad i\ne j\\
\Gamma^i_{ij}&:=\f{H_j}{H_i}\beta_{ij}\qquad i\ne j\\
\Gamma^i_{ii}&:=-\sum_{l\ne i}\f{u^l}{u^i}\Gamma^i_{li}-\f{1}{u^i}
\end{split}
\end{equation}
 is flat.
\end{theorem}

\noindent
\emph{Proof}.  For distinct indices clearly the components of the Riemann tensor vanish. For $i\ne j\ne k\ne i$:
\begin{eqnarray*}
&&R^i_{jki}=-R^i_{jik}=\d_k\Gamma^i_{ij}+\Gamma^i_{ik}\Gamma^i_{ij}-\Gamma^i_{ij}\Gamma^j_{jk}
-\Gamma^i_{ik}\Gamma^k_{kj}=\\
&&\d_k\left(\f{H_j}{H_i}\beta_{ij}\right)+\f{H_k H_j}{(H_i)^2}\beta_{ik}\beta_{ij}-\f{H_k}{H_i}\beta_{ij}\beta_{jk}
-\f{H_j}{H_i}\beta_{ik}\beta_{kj}=\\
&&\f{\d_k H_j}{H_i}\beta_{ij}-\f{H_j\d_k H_i}{(H_i)^2}\beta_{ij}+\f{H_j}{H_i}\beta_{ik}\beta_{kj}+\f{H_k H_j}{(H_i)^2}\beta_{ik}\beta_{ij}-\f{H_k}{H_i}\beta_{ij}\beta_{jk}
-\f{H_j}{H_i}\beta_{ik}\beta_{kj}=0.
\end{eqnarray*}
The vanishing of
\begin{equation*}
R^i_{jjk}=-R^i_{jkj}=-\d_k\Gamma^i_{jj}
+\Gamma^i_{jj}\Gamma^j_{jk}
-\Gamma^i_{ik}\Gamma^i_{jj}
-\Gamma^i_{kk}\Gamma^k_{jj}=0\,\,\,\mbox{if $i\ne k\ne j\ne i$}
\end{equation*}
follows from the vanishing of $R^i_{jki}$ and from 
 $\Gamma^i_{jj}=-\f{u^i}{u^j}\Gamma^i_{ij}$. Indeed using the last identity we obtain
\begin{eqnarray*}
R^i_{jjk}&=&\f{u^i}{u^j}\d_k\Gamma^i_{ij}
-\f{u^i}{u^j}\Gamma^i_{ij}\Gamma^j_{jk}
+\f{u^i}{u^j}\Gamma^i_{ik}\Gamma^i_{ij}
-\f{u^i}{u^k}\f{u^k}{u^j}\Gamma^i_{ik}\Gamma^k_{kj}\\
&=&\f{u^i}{u^j}\left[\d_k\Gamma^i_{ij}
-\Gamma^i_{ij}\Gamma^j_{jk}
+\Gamma^i_{ik}\Gamma^i_{ij}
-\Gamma^i_{ik}\Gamma^k_{kj}\right]=\f{u^i}{u^j}R^i_{jki}=0.
\end{eqnarray*}  
The vanishing of  
$$R^i_{ikl}=-R^i_{ilk}=\d_k\Gamma^i_{il}-\d_l\Gamma^i_{ik}=0\,\,\,\mbox{if $i\ne k\ne l\ne i$}$$
is immediate using the vanishing of $R^i_{jki}$ and the contracted first Bianchi identity $R^i_{ijk}+R^i_{jki}+R^i_{kij}=0.$

Finally, for $i\ne j$ we have:
\begin{eqnarray*}
&&R^i_{iji}=-R^i_{iij}=\d_j\Gamma^i_{ii}-\d_i\Gamma^i_{ij}+\Gamma^j_{ii}\Gamma^i_{jj}-\Gamma^j_{ij}\Gamma^i_{ji}=\\
&&\d_j\Gamma^i_{ii}-\d_i\Gamma^i_{ij}=
-\sum_{l\ne i}\f{u^l}{u^i}\d_j\Gamma^i_{il}-\f{1}{u^i}\f{H_j}{H_i}\beta_{ij}-\d_i\left(\f{H_j}{H_i}\beta_{ij}\right)=\\
&&-\sum_{l\ne i,j}\f{u^l}{u^i}\f{\d_j H_l}{H_i}\beta_{il}
+\sum_{l\ne i,j}\f{u^l}{u^i}\f{H_l\d_j H_i}{(H_i)^2}\beta_{il}-\sum_{l\ne i,j}\f{u^l}{u^i}\f{H_l}{H_i}\d_j\beta_{il}
+\f{u^j}{u^i}\f{H_j\d_j H_i}{(H_i)^2}\beta_{ij}+\\
&&+\f{1}{u^iH_i}\left(\sum_{l\ne j}u^l\d_l H_j+dH_j\right)\beta_{ij}
-\f{1}{u^i}\f{H_j}{H_i}\beta_{ij}
-\f{\d_i H_j}{H_i}\beta_{ij}-\f{H_j}{u^i(H_i)^2}\left(\sum_{l\ne i}u^l\d_l H_i+dH_i\right)\beta_{ij}+\\
&&-\f{H_j}{H_i}\f{u^j\d_j\beta_{ij}+u^i\d_i\beta_{ij}}{u^i}=\\
&&-\sum_{l\ne i,j}\f{u^l}{u^i}\f{\d_j H_l}{H_i}\beta_{il}
+\sum_{l\ne i,j}\f{u^l}{u^i}\f{H_l\d_j H_i}{(H_i)^2}\beta_{il}-\sum_{l\ne i,j}\f{u^l}{u^i}\f{H_l}{H_i}\beta_{ij}\beta_{jl}
+\f{1}{u^iH_i}\left(\sum_{l\ne i,j}u^l\d_l H_j\right)\beta_{ij}\\
&&-\f{H_j}{u^i(H_i)^2}\left(\sum_{l\ne i,j}u^l\d_l H_i\right)\beta_{ij}+\f{H_j}{u^iH_i}\sum_{l\ne i,j}u^l\d_l\beta_{ij}=0,
\end{eqnarray*}
and
\begin{eqnarray*}
&&R^i_{jji}=-R^i_{jij}=\d_j\Gamma^i_{ji}-\d_i\Gamma^i_{jj}+(\Gamma^i_{ij})^2+\Gamma^i_{jj}\Gamma^j_{ji}
-\sum_{l=1}^n\Gamma^i_{li}\Gamma^l_{jj}=\\
&&\f{u^j\d_j\Gamma^i_{ij}+u^i\d_i\Gamma^i_{ij}}{u^j}+\f{\Gamma^i_{ij}}{u^j}+\Gamma^i_{ij}\left(\Gamma^i_{ij}-\f{u^i}{u^j}\Gamma^j_{ji}\right)
-\sum_{l\ne i,j}\f{u^l}{u^j}\Gamma^i_{il}\Gamma^i_{ij}+\\
&&-(\Gamma^i_{ij})^2+\sum_{l\ne i,j}\f{u^l}{u^j}\Gamma^i_{ij}\Gamma^j_{jl}
+\f{u^i}{u^j}\Gamma^i_{ij}\Gamma^j_{ji}+\sum_{l\ne i,j}\f{u^l}{u^j}\Gamma^i_{il}\Gamma^l_{lj}=\\
&&-\f{H_j}{u^j H_i}\left[\sum_{l\ne i,j}u^l\d_l\beta_{ij}+\beta_{ij}\right]+\f{\beta_{ij}}{u^j H_i}(u^i\d_i+u^j\d_j)H_j
-\f{H_j\beta_{ij}}{u^j(H_i)^2}(u^i\d_i+u^j\d_j)H_i+\f{\Gamma^i_{ij}}{u^j}\\
&&-\sum_{l\neq i,j}\frac{u^l}{u^j}\Gamma^{i}_{il}\Gamma^i_{ij}+\sum_{l\ne i,j}\f{u^l}{u^j}\Gamma^i_{ij}\Gamma^j_{jl}
+\sum_{l\ne i,j}\f{u^l}{u^j}\Gamma^i_{il}\Gamma^l_{lj}=\\
&&-\sum_{l\ne i,j}\f{u^l}{u^j}\f{H_j}{H_i}\beta_{il}\beta_{lj}-\left(\sum_{l\ne i,j}\f{u^l}{u^j}\f{\d_l H_j}{H_l}\f{H_l}{H_i}
\beta_{ij}+\f{\beta_{ij}H_j d}{u^j H_i}\right)+\left(\sum_{l\ne i,j}\f{u^l}{u^j}\f{H_jH_l}{(H_i)^2}\f{\d_l H_i}{H_l}\beta_{ij}+\f{H_j\beta_{ij} d}{u^jH_i}\right)
\\&&-\sum_{l\ne i,j}\f{u^l}{u^j}\f{H_lH_j}{(H_i)^2}\beta_{il}
\beta_{ij}+\sum_{l\ne i,j}\f{u^l}{u^j}\f{H_l}{H_i}\beta_{ij}\beta_{jl}
+\sum_{l\ne i,j}\f{u^l}{u^j}\f{H_j}{H_i}\beta_{il}\beta_{lj}=0.
\end{eqnarray*}
This proves the claim.
\endproof

\begin{defi}
We call $\nabla_2$ the \emph{dual connection} associated with $(\beta_{ij},H_i)$.
\end{defi}

\begin{lemma}\label{lemmad}
The connection $\nabla_2$ defined in \eqref{dualnabla} is compatible with the product $c^{*i}_{jk}=\f{\delta^i_j\delta^i_k}{u^i}$
 and satisfies $\nabla_2 E=0$. 
\end{lemma}
\proof
The fact that $\nabla_2 E=0$ identically is equivalent to the requirements $\Gamma^i_{jj}:=-\f{u^i}{u^j}\Gamma^i_{ij}$ for $i\ne j$
 and $\Gamma^i_{ii}:=-\sum_{l\ne i}\f{u^l}{u^i}\Gamma^i_{li}-\f{1}{u^i}$ that appear in \eqref{dualnabla}.

Moreover, the compatibility of $\nabla_2$  with $c^{*i}_{jk}:=\f{\delta^i_j\delta^i_k}{u^i}$ is equivalent to
 the requirements $\Gamma^i_{jj}:=-\f{u^i}{u^j}\Gamma^i_{ij}$ for $i\ne j$
  and $\Gamma^i_{jk}=0$ for $i\ne j\ne k\ne i$. All  these statements can be checked via straightforward
 computations in coordinates $\{u^1, \dots, u^n\}$. The Lemma is proved.
\endproof


 


Summarizing, combining Theorem \ref{dualnablaflathm} and Lemma \ref{lemmad} we have proved the following:
\begin{theorem}\label{mth2}
Let $(\beta_{ij},H_i)$ be a solution of the system  (\ref{ED1},\ref{ED3}) and (\ref{L1},\ref{L3}), then 
 the dual connection $\nabla_2$ associated with $(\beta_{ij},H_i)$ and the structure constants  defined
 in the coordinates $(u^1,\dots,u^n)$ by $c^{*i}_{jk}=\f{1}{u^i}\delta^i_j\delta^i_k$
 give rise to an $F$-manifold with compatible  flat structure. 
\end{theorem}

\section{Bi-flat $F$-manifolds}
Given a solution of the system (\ref{L1},\ref{L2},\ref{L3}), applying the results of the previous sections,
 we can construct a bi-flat $F$-manifold according to Definition \ref{defibiflat}. Indeed the natural connection and the dual connection
 associated to the same functions $H_i$ are almost hydrodymically equivalent by definition since 
$$\Gamma^{(1)i}_{ij}=\Gamma^{(2)i}_{ij}=\f{H_j}{H_i}\beta_{ij}.$$
 The problem is that, for arbitary values of the constant $d$, system (\ref{L1},\ref{L2},\ref{L3}) 
  does not admit solutions. The choice of the right degree of homogeneity can be done following the same 
 procedure used by Dubrovin in \cite{du93} for the symmetric case.
 The key point is the observation that homogeneous solutions
 of the system (\ref{L1},\ref{L2}) are the eigenvectors of a matrix $V$ of components $V_{ij}=(u^j-u^i)\beta_{ij}$.
 It turns out that the eigenvalues of $V$ are the admissible degrees of homogeneity
 of the solutions of the system (\ref{L1},\ref{L2}).
 The only difference with respect to the symmetric case is that,  
 in the general case, the matrix $V$ is not skew-symmetric. We refer the reader to \cite{du93} (and in particular
 to lemma 3.9 and corollary 3.2) for all details.

Combining theorems (\ref{mth1}) and (\ref{mth2}) we obtain
\begin{theorem}\label{mainth}
Let $\beta_{ij}$ be a solution of the system  (\ref{ED1},\ref{ED2},\ref{ED3}) and $H_i$
 a homogeneous solution of the system (\ref{L1},\ref{L2}), then  
\begin{itemize}
\item the natural connection $\nabla_1$ associated with $(\beta_{ij},H_i)$, 
\item the dual connection $\nabla_2$ associated with $(\beta_{ij},H_i)$,
\item the structure constants  defined in the coordinates $(u^1,\dots,u^n)$ by $c^i_{jk}=\delta^i_j\delta^i_k$,
\item the structure constants  defined in the coordinates $(u^1,\dots,u^n)$ by $c^{*i}_{jk}=\f{1}{u^i}\delta^i_j\delta^i_k$, 
\item the vector fields $e=\sum_{i=1}^n\f{\d}{\d u^i}$ and $E=\sum_{i=1}^n u^i\f{\d}{\d u^i}$,  
\end{itemize}
define a bi-flat semisimple $F$-manifold $(M,\nabla_1,\nabla_2,\circ,*,e,E)$.
\end{theorem}

\begin{remark}
In the generic case, the matrix $V$ has $n$ distinct eigenvalues and thus
 there are $n$ possible choices for the natural connection,
 one for each independent homogeneous solution $H^{(\alpha)}_i,\alpha=1,\dots,n$ of the system (\ref{L1},\ref{L2}).
  The flat coordinates $(t^1,\dots,t^n)$ of such connections satify a system involving the 
specified solution $H^{(\alpha)}_i$ and a basis 
 of solution $K^{(\beta)}_i,\beta=1,\dots,n$ of the adjoint system
\begin{eqnarray}
\label{L1ad}
&&\d_j K_i=\beta_{ji}K_j,\qquad  i\ne j\\
\label{L2ad}
&&e(K_i)=0.
\end{eqnarray}
More precisely we have
\beq\label{ps}
\d_i t^{\beta}=K_i^{(\beta)}H_i^{(\alpha)}
\eeq
The proof is easy. First one observes that the system \eqref{ps} is compatible. Then, by straightforward
 computation it is easy to check that the 1-forms $\omega^{\beta}$ defined as $\omega^{\beta}:=\sum_{i=1}^n(K_i^{(\beta)}H_i^{(\alpha)})\,du^i$
 are flat. Indeed for $j\ne i$ we have:
\begin{eqnarray*}
\nabla_j\omega^{\beta}_i&=&\d_j\omega^{\beta}_i-\Gamma^i_{ij}\omega^{\beta}_i-\Gamma^j_{ji}\omega^{\beta}_j=\\
&&=\d_j(K^{(\beta)}_i H^{(\alpha)}_i)-\f{H^{(\alpha)}_j}{H^{(\alpha)}_i}\beta_{ij}K^{(\beta)}_iH^{(\alpha)}_i-
\f{H^{(\alpha)}_i}{H^{(\alpha)}_j}\beta_{ji}K^{(\beta)}_j H^{(\alpha)}_j=0.
\end{eqnarray*}
While for $j=i$ we obtain:
\begin{eqnarray*}
\nabla_i\omega^{\beta}_i&=&\d_i\omega^{\beta}_i-\Gamma^l_{ii}\omega^{\beta}_l=
 \d_i\omega^{\beta}_i-\Gamma^i_{ii}\omega^{\beta}_i-\sum_{l\ne i}\Gamma^l_{ii}\omega^{\beta}_l
=\d_i\omega^{\beta}_i+\sum_{l\ne i}\Gamma^i_{il}\omega^{\beta}_i+\sum_{l\ne i}\Gamma^l_{li}\omega^{\beta}_l=\\
&&\d_i(K^{(\beta)}_i H^{(\alpha)}_i)+\sum_{l\ne i}\f{H^{(\alpha)}_l}{H^{(\alpha)}_i}\beta_{il}K_i^{(\beta)}H_i^{(\alpha)}
+\sum_{l\ne i}\f{H^{(\alpha)}_i}{H^{(\alpha)}_l}\beta_{li}K_l^{(\beta)}H_l^{(\alpha)}=\\
&&\sum_{i\ne l}\left(-\beta_{li}K^{(\beta)}_l H^{(\alpha)}_i-\beta_{il}K^{(\beta)}_i H^{(\alpha)}_l
+\beta_{il}K^{(\beta)}_i H^{(\alpha)}_l+\beta_{li}K^{(\beta)}_l H^{(\alpha)}_i\right)=0.
\end{eqnarray*}
System \eqref{ps} appears also in the recent paper \cite{ps} devoted to oriented associativity equations. The reason
 is that the structure constants of an $F$-manifold with compatible flat structure admit, in flat coordinates,
 a vector potential and, as a consequence, the associativity equations in flat coordinates reduce to oriented associativity
 equations.
\end{remark}


\section{Examples in the case $n=2$}
The aim of this section is to find solutions of the system (\ref{ED1},\ref{ED2},\ref{ED3}) and to construct the
 corresponding $F$-manifolds with compatible flat structure.

\subsection{Egorov-Darboux system}
In this case the Egorov-Darboux system reduces to
\begin{eqnarray*}
&&\f{\d\beta_{ij}}{\d u^1}+\f{\d\beta_{ij}}{\d u^2}=0,\\
&&u^1\f{\d\beta_{ij}}{\d u^1}+u^2\f{\d\beta_{ij}}{\d u^2}=-\beta_{ij}.
\end{eqnarray*}
The first equations tell us that the rotation coefficients depend only on the difference $(u^1-u^2)$. The remaining equations tell us that they
 are homogeneous
 functions of degree $-1$. This gives us
\begin{eqnarray*}
\beta_{12}&=&\f{C_1}{u^1-u^2},\\
\beta_{21}&=&\f{C_2}{u^1-u^2}.
\end{eqnarray*}
\subsection{Natural connections}
To construct the natural connections we need to compute
 the Lam\'e coefficients, i.e. we solve the system:
\begin{eqnarray*}
&&\f{\d H_i}{\d u^1}+\f{\d H_i}{\d u^2}=0,\\
&&\d_2 H_1=\f{C_1}{u^1-u^2}H_2,\\
&&\d_1 H_2=\f{C_2}{u^1-u^2}H_1.
\end{eqnarray*}
Again, the first equation tells us that the Lam\'e coefficients depend only on the difference $z=u^1-u^2$.
 The remaining equations are equivalent to
 the system:
\begin{eqnarray*}
&&\d^2_z H_{1}+\f{1}{z}\d_z H_1+\f{C_1 C_2}{z^2}H_1=0,\\
&&H_2=-\f{z}{C_1}\d_z H_1,
\end{eqnarray*}
that gives us
\begin{eqnarray}
H_1 &=&a\sin \left( \sqrt {C_1C_2}\ln{(u^1-u^2)}\right) +b\,\cos
 \left( \sqrt{C_1C_2}\ln{(u^1-u^2)}
 \right), \\  
H_2&=&-\sqrt {\f{C_2}{C_1}}\left[a\cos \left( \sqrt {C_1C_2}\ln{(u^1-u^2)}\right) -b\sin
 \left( \sqrt {C_1C_2}\ln{(u^1-u^2)} 
 \right)  \right],
\end{eqnarray}
where $a$ and $b$ are arbitrary constants. Then by definition, the natural connection $\nabla^{(1)}$ is given by
\begin{eqnarray*}
&&\Gamma^1_{11}=\Gamma^1_{22}=-\Gamma^1_{12}=-\Gamma^1_{21},\\
&&                                                                        \\
&&\Gamma^1_{12}={\frac {\sqrt{C_1 C_2}\left( -a\cos
 \left( \sqrt {C_1C_2}\ln{(u^1-u^2)} 
 \right) +b\sin \left( \sqrt{C_1C_2}\ln{(u^1-u^2)}\right)\right) }{ \left( a\sin \left( 
\sqrt {C_1C_2}\ln{(u^1-u^2)}\right) +b\cos \left( \sqrt{C_1C_2}\ln{(u^1-u^2)}\right)\right)(u^1-u^2)}}, \\
&&                                                                           \\
&&\Gamma^2_{22}=\Gamma^2_{11}=-\Gamma^2_{21}=-\Gamma^2_{12},\\
&&                                                                                                   \\
&&\Gamma^2_{21}=\f{\sqrt{C_1C_2}\left(a\sin \left( \sqrt {C_1C_2}\ln{(u^1-u^2)}\right) +b\,\cos
 \left( \sqrt{C_1C_2}\ln{(u^1-u^2)}
 \right)\right)}{\left(-a\cos \left( \sqrt {C_1C_2}\ln{(u^1-u^2)}\right) +b\sin
 \left( \sqrt {C_1C_2}\ln{(u^1-u^2)} 
 \right)\right)(u^1-u^2)}.
\end{eqnarray*}
\subsection{Dual connections}
Let us consider now the dual connections. They are defined starting from the Lam\'e coefficients $H_1$ and $H_2$
 satisfying the system:
\begin{eqnarray*}
&&u^1\f{\d H_i}{\d u^1}+u^2\f{\d H_i}{\d u^2}=dH_i,\\
&&\d_2 H_1=\f{C_1}{u^1-u^2}H_2,\\
&&\d_1 H_2=\f{C_2}{u^1-u^2}H_1.
\end{eqnarray*}
The last two equations can be written as a second order linear equation for $H_1(u^1,u^2)$. Indeed, since
$$H_2(u^1,u^2)=\f{(u^1-u^2)\d_2 H_1}{C_1},$$
we obtain
$$(u^1-u^2)^2\f{\d^2 H_1}{\d u^1\d u^2}+(u^1-u^2)\f{\d H_1}{\d u^2}-C_1 C_2H_1=0.$$
Taking into account the first equation we have
$$H_1(u^1,u^2)=f\left(\f{u^2}{u^1}\right)(u^1)^d$$
and the second order PDE for $H_1$ reduces to the following second order ODE for $f(z)$ ($z:=\f{u^2}{u^1}$):
\beq\label{IIorderODE}
\left( {\frac {d^{2}}{d{z}^{2}}}f \left( z \right)  \right) z \left( 
z-1 \right) ^{2}+ \left( {\frac {d}{dz}}f \left( z \right)  \right) 
 \left( {z}^{2}-z-d \left( z-1 \right) ^{2} \right) +C_1C_2f(z),
\eeq
which is very similar to Euler's hypergeometric differential equation. It is not surprising then that
the general solution of \eqref{IIorderODE} can be written in terms of Gauss' hypergeometric functions $\phantom{}_2F_1(a,b;c;z)$
 (for more information about Gauss' hypergeometric functions see for instance \cite{WW});
 indeed it turns out that the general solution in a neighborhood of $z=0$ can be written as
 $$f \left( z \right) ={ a}\, \left( z-1 \right) ^{-i\sqrt {{
C_1}}\sqrt {{C_2}}}
{\phantom{}_2F_1(-i\sqrt {{ C_1}}\sqrt {{ C_2}},-i\sqrt {{ C_1}}\sqrt {{C_2}}-d;\,-d;\,z)}
+$$
$$+{b}\,{z}^{d+1} \left( z-1 \right) ^{-i\sqrt {{ C_1}}\sqrt 
{{ C_2}}}
{\phantom{}_2F_1(-i\sqrt {{ C_1}}\sqrt {{C_2}}+1,-i\sqrt {{ C_1}}\sqrt {{ C_2}}+d+1;\,d+2;\,z)},
$$ where $a$ $b$ are arbitrary constants of integration. For special values of the parameters $C_1$ and $C_2$
 the hypergeometric functions $\phantom{}_2F_1$ reduce to well known elementary functions. 

Let us consider, for instance, the particular case
 correponding to $C_1=1$ and $C_2=-4$. In this special case the general solution of \eqref{IIorderODE} 
is given by
$$f \left( z \right) =a\,{\frac {\left( -z+dz-2-d \right) {z}^
{d+1}}{ \left( z-1 \right) ^{2}}}+b\,{\frac {  \left( 
{d}^{2}+3\,d+2 \right) {z}^{2}+ \left( -2\,d-2\,{d}^{2}+4 \right) z+{d
}^{2}-d  }{ \left( z-1 \right) ^{2}}},$$
where $a$ and $b$ are arbitrary constants. Summarizing we obtain
\begin{eqnarray*}
H_1&=&-a(u^2)^{d+1}\left[{\frac {u^2-du^2+2u^1+du^1}{ \left(u^1-u^2\right)^{2}}}\right]+\\
&&b(u^1)^{d}\left[{\frac {
  (u^2)^{2}({d}^{2}+3d+2)+
u^1u^2(-2d-2d^2+4)+(u^1)^{2}(d^2-d)}
{ \left(u^1-u^2 \right) ^{2}}}\right]\\
H_2&=&-4b(u^1)^{d+1} \left[{\frac{-du^2+du^1-u^1-2u^2 }{ \left(u^1-u^2\right) ^{2}}}\right]\\
&&-a(u^2)^d\left[{\frac {(u^2)^{2}(d^{2}-d)+u^1u^2(-2{d}^{2}-2d+4)+(u^1)^{2}(2+3d+{d}^{2})}{\left(u^1-u^2 \right) ^{2}}}\right]
\end{eqnarray*}
\subsection{Bi-flat $F$-manifolds}
To construct bi-flat $F$-manifolds in the case $n=2$ we have to solve the following over-determined system:
\begin{eqnarray*}
&&\f{\d H_i}{\d u^1}+\f{\d H_i}{\d u^2}=0,\\
&&u^1\f{\d H_i}{\d u^1}+u^2\f{\d H_i}{\d u^2}=dH_i,\\
&&\d_2 H_1=\f{C_1}{u^1-u^2}H_2,\\
&&\d_1 H_2=\f{C_2}{u^1-u^2}H_1.
\end{eqnarray*}
It is easy to check that solutions are given by the formulas:
\begin{eqnarray*}
H_1&=&D_1(u^1-u^2)^d,\\
H_2&=&D_2(u^1-u^2)^d.
\end{eqnarray*}
where the constants $D_1, D_2$ and $d$ obbey two additional additional constraints:
$$-C_1\f{D_2}{D_1}=C_2\f{D_1}{D_2}=d.$$
Multiplying both the constraints we obtain
$$d^2=-C_1 C_2.$$
The same result can be obtained computing the eigenvalues and the eigenvector of the matrix $V$ that in this
 case reads
\beq
\begin{pmatrix}
0 & -C_1\\
C_2 & 0
\end{pmatrix}.
\eeq
 For any choice of $C_1$ and $C_2$ 
the natural and dual connections $\nabla_1$ and $\nabla_2$ are defined by \eqref{naturalnabla} and \eqref{dualnabla} with
\begin{eqnarray*}
\Gamma^1_{12}&=&\f{D_2}{D_1}\f{C_1}{u^1-u^2}=\f{-d}{u^2-u^1}\\
\Gamma^2_{21}&=&\f{D_1}{D_2}\f{C_2}{u^1-u^2}=\f{d}{u^2-u^1}.
\end{eqnarray*}

\section{Bi-flat $F$-manifolds in dimension $n=3$}
In this Section, we study the Darboux-Egorov system augmented with the condition $E(\beta_{ij})=-\beta_{ij}$ in the case in which $n=3$. 
 This system was studied by several authors \cite{K,CGM,KK,AVdL}. However as far as we know, in literature there are no explicit formulas to
 obtain the solutions of (\ref{ED1},\ref{ED2}) starting from  solutions of Painlev\'e VI.
  The proof of the equivalence we present here is completely elementary and part of the proof concerning the reduction to Painlev\'e VI
 is drawn on \cite{AVdL}.
The main Theorem of this section shows that bi-flat $F$-manifolds in dimension $n=3$ are parametrized by solutions of a two-parameter Painlev\'e VI equation. 

First we show that the augmented Darboux-Egorov system is equivalent to a system of non-autonomous ODEs. 
\begin{proposition}
In dimension $n=3$, on the open set $u^1\neq u^2\neq u^3\neq u^1$, 
the system (\ref{ED1}, \ref{ED2}, \ref{ED3}) 
 is equivalent to the following non-autonomous system of ODEs:
\begin{equation}\label{reductionODE3}
\begin{split}
  \frac{d}{dz} F_{12}(z)=\frac{1}{z(z-1)}F_{13}(z)F_{32}(z)\\
 \frac{d}{dz} F_{13}(z)=-\frac{1}{z-1}F_{12}(z)F_{23}(z)\\
\frac{d}{dz} F_{21}(z)=\frac{1}{z(z-1)}F_{23}(z)F_{31}(z) \\
 \frac{d}{dz} F_{23}(z)=\frac{1}{z}F_{21}(z)F_{13}(z)\\
\frac{d}{dz} F_{31}(z)=-\frac{1}{z-1}F_{32}(z)F_{21}(z) \\
\frac{d}{dz} F_{32}(z)=\frac{1}{z}F_{31}(z)F_{12}(z),\\
\end{split}
\end{equation}
where the independent variable $z:=\frac{u^3-u^1}{u^2-u^1}$.  The unknown functions $\beta_{ij}$ are given in terms of the solutions $F_{ij}$ of the system above as follows:
\begin{equation}\label{relationsbetaF}
\begin{split}
\beta_{12}(u^1, u^2, u^3)=\frac{1}{u^2-u^1}F_{12}\left(\frac{u^3-u^1}{u^2-u^1} \right)\\
\beta_{21}(u^1, u^2, u^3)=\frac{1}{u^2-u^1}F_{21}\left(\frac{u^3-u^1}{u^2-u^1} \right)\\
\beta_{32}(u^1, u^2, u^3)=\frac{1}{u^3-u^2}F_{32}\left(\frac{u^3-u^1}{u^2-u^1} \right)\\
\beta_{23}(u^1, u^2, u^3)=\frac{1}{u^3-u^2}F_{23}\left(\frac{u^3-u^1}{u^2-u^1} \right)\\
\beta_{13}(u^1, u^2, u^3)=\frac{1}{u^3-u^1}F_{13}\left(\frac{u^3-u^1}{u^2-u^1} \right)\\
\beta_{31}(u^1, u^2, u^3)=\frac{1}{u^3-u^1}F_{31}\left(\frac{u^3-u^1}{u^2-u^1} \right)\\
\end{split}
\end{equation}

\end{proposition}
\proof
The equations $e(\beta_{ij})=0$ are equivalent to the requirement that for each pair $(i,j)$, $i\neq j$, $\beta_{ij}$ is an arbitrary function $G_{ij}$ of the differences of the coordinates. Therefore we can write $\beta_{ij}(u^1, u^2, u^3):=G_{ij}(x_{31}, x_{21})$ where $x_{31}:=u^3-u^1$ and $x_{21}:=u^2-u^1$ and with these, the subsystem $e(\beta_{ij})=0$ is automatically satisfied. If we substitute the unknown functions $G_{ij}(x_{31}, x_{21})$ in the subsystem $E(\beta_{ij})=-\beta_{ij}$, we obtain the following equations for each $i,j$, $i\neq j$: 
\beq\label{eulerobeta3}\frac{\partial}{\partial x_{31}}G_{ij}(x_{31}, x_{21})+\frac{\partial}{\partial x_{21}}G_{ij}(x_{31}, x_{21})+G_{ij}(x_{31}, x_{21})=0.\eeq

It turns out that each equation of \eqref{eulerobeta3} is equivalent to require that $G_{ij}$ is an arbitrary function of the {\em ratio} $\frac{x_{31}}{x_{21}}$ divided by $x_{21}$. So if we set 
$$G_{ij}(x_{31}, x_{21}):=\frac{1}{x_{21}}K_{ij}\left(\frac{x_{31}}{x_{21}} \right),$$
then also the subsystem $E(\beta_{ij})=-\beta_{ij}$ is identically satisfied. Let us observe that this equivalence holds only whenever $x_{21}\neq 0$, i.e. when $u^2\neq u^1$. However, since equation \eqref{eulerobeta3} is symmetric with respect to the exchange of $x_{21}$ and $x_{31}$ we could have analogously found a solution $\tilde K_{ij}=\tilde K_{ij}\left(\frac{x_{21}}{x_{31}} \right)$. Therefore, for the equivalence to hold without analyzing different cases, we impose also that $u^3\neq u^1$.  

Now it remains to express the subsystem $\partial_j\beta_{ik}=\beta_{ij}\beta_{jk}$ in terms of the functions $K_{ij}$. This turns out to be possible, in particular defining the variable $z:=\frac{x_{31}}{x_{21}}=\frac{u^3-u^1}{u^2-u^1}$ it is an easy computation to show that in dimension $3$, the system $\partial_j\beta_{ik}=\beta_{ij}\beta_{jk}$ is equivalent to the following system of ODEs:
\begin{equation}
\begin{split}
{\frac {d}{dz}}{K_{12}} \left( z \right) -{K_{13}} \left( z \right) 
{K_{32}} \left( z \right) =0 \\
-z{\frac {d}{dz}}{K_{13}} \left( z \right) -{K_{13}} \left( z
 \right) -{K_{12}} \left( z \right) {K_{23}} \left( z \right)  = 0\\
{\frac {d}{dz}}{K_{21}} \left( z \right) -{K_{23}} \left( z \right) 
{K_{31}} \left( z \right) = 0\\
\left( z-1 \right) {\frac {d}{dz}}{K_{23}} \left( z \right) +{K_{23}} \left( z \right) -{K_{21}} \left( z \right) {K_{13}} \left( z
 \right)=0 \\
-z{\frac {d}{dz}}{K_{31}} \left( z \right) -{K_{31}} \left( z
 \right) -{K_{32}} \left( z \right) {K_{21}} \left( z \right)=  0\\
 \left( z-1 \right) {\frac {d}{dz}}{K_{32}} \left( z \right) +{K_{32}} \left( z \right) -{K_{31}} \left( z \right) {K_{12}} \left( z
 \right) =  0 \\
 \end{split}
 \end{equation}
 
 Performing a further change of variables, namely defining $ F_{32}(z):=(z-1)K_{32}(z)$, $F_{23}(z):=(z-1)K_{23}(z)$, $F_{13}(z):=z K_{13}(z)$, and $F_{31}(z):=zK_{31}(z)$, $F_{12}(z):=K_{12}(z)$ and $F_{21}(z):=K_{21}(z)$ the system above is transformed to the following one: 
 
 \begin{equation}\label{finalODE3}
 \begin{split}
  \frac{d}{dz} F_{12}(z)=\frac{1}{z(z-1)}F_{13}(z)F_{32}(z)\\
\frac{d}{dz} F_{13}(z) = -\frac{1}{z-1}F_{12}(z)F_{23}(z)\\
\frac{d}{dz} F_{21}(z) = \frac{1}{z(z-1)}F_{23}(z)F_{31}(z) \\
 \frac{d}{dz} F_{23}(z)= \frac{1}{z}F_{21}(z)F_{13}(z)\\
\frac{d}{dz} F_{31}(z) =  -\frac{1}{z-1}F_{32}(z)F_{21}(z) \\
\frac{d}{dz} F_{32}(z) = \frac{1}{z}F_{31}(z)F_{12}(z).
\end{split}
\end{equation}
Observe that in order to write this system in normal form, namely with the derivatives having coefficients equal to $1$ we must divide by $z-1$, which means we must impose the also the condition $u^3\neq u^2$. 

Finally the expressions relating  $\beta_{ij}$ to $F_{ij}$ are obtained through a simple computation. 
\endproof

Let us observe that the non-autonomous system of ODEs for the $F_{ij}$  reduces to the Hamiltonian system on $\mathfrak{so}(3)$ given in (\cite{du93}, Lecture 3, $(3.113)$) if we consider the reduction $F_{ij}=F_{ji}$. 

Now we discuss how the non-autonomous systems of ODEs for the $F_{ij}$ is related in this case to a Painlev\'e VI equation with two independent parameters. 

\begin{theorem}\label{painleveVI}
System \eqref{finalODE3} is equivalent to the following  sigma form of Painlev\'e VI equation:
\begin{equation}\label{sigmapainleve0}
\begin{split}
z^2(z-1)^2\left(\frac{d^2\sigma}{dz^2}\right)^2+4\left[\frac{d\sigma}{dz}\left(z\frac{d\sigma}{dz}-\sigma\right)-\left(\frac{d\sigma}{dz}\right)^2\left(z\frac{d\sigma}{dz}-\sigma\right) \right]=\\
\left(\frac{d\sigma}{dz}\right)^2\left(v_1^2+v_2^2+v_3^2 \right)+\frac{d\sigma}{dz}\left(v_1^2v_2^3+v_1^2v_3^2+v_2^2v_3^2 \right)+v_1^2v_2^2v_3^2, 
\end{split}
\end{equation}
where the three parameters $v_1, v_2, v_3$ are the three roots of the polynomial 
$$x^3-2R^2x^2+R^4x-D^2,$$
where the constants $-R^2$ and $D$ are expressed in terms of the conserved quantities of system \eqref{finalODE3} given by: 
$$F_{12}(z)F_{21}(z)+F_{13}(z)F_{31}(z)+F_{23}(z)F_{32}(z)=-R^2,$$
$$F_{23}(z)F_{31}(z)F_{12}(z)-F_{13}(z)F_{32}(z)F_{21}(z)=D.$$
In particular system \eqref{finalODE3} is equivalent to a Painlev\'e VI depending on two parameters, since $v_i$ are expressed in terms of the quantities $-R^2$ and $D$, $v_i=v_i(-R^2, D)$. 
\end{theorem}
\proof
First notice that 
$$\frac{d}{dz}\left(F_{12}(z)F_{21}(z)+F_{13}(z)F_{31}(z)+F_{23}(z)F_{32}(z)\right)=0,$$
identically along the solutions of \eqref{finalODE3}, as a simple computation shows. So we set
\begin{equation}\label{constant1}
F_{12}(z)F_{21}(z)+F_{13}(z)F_{31}(z)+F_{23}(z)F_{32}(z)=-R^2, 
\end{equation}
where $R$ is a not necessarily real constant. The choice of the minus sign and the square is dictated by the fact that it will be easier to express the roots of a cubic polynomial in terms of its coefficient with this choice and these roots identify the parameters in the sigma form of Painlev\'e VI.
Since $F_{ij}$ are all functions of $z$, we can always find a function $f(z)$ such that \begin{equation}\label{position12}F_{12}(z)F_{21}(z):=f'(z)\end{equation}
identically, where $f'$ denotes the derivative of $f$ with respect to $z$. Notice that $f$ is determined up to a constant. We want to express $F_{13}(z)F_{31}(z)$ also in terms of $f$, $f'$ and $z$. Due to equations \eqref{finalODE3}, we have $\frac{d}{dz}\left(F_{13}F_{31} \right)=-z\frac{d}{dz}\left(F_{12}F_{21}\right)$ and therefore
$$\frac{d}{dz}\left(F_{13}F_{31} \right)= -z\frac{d}{dz}\left(F_{12}F_{21}\right) =F_{12}F_{21}-\frac{d}{dz}\left(zF_{12}F_{21} \right).$$
Substituting the expression of $F_{12}F_{21}$ in terms of $f'$ and integrating with respect to $z$ we find 
\begin{equation}\label{position13}
F_{13}(z)F_{31}(z):=f(z)-zf'(z)-R^2,
\end{equation}
where part of the constant of integration has been absorbed in $f$ in order to explicit $R$, the constant appearing in \eqref{constant1}. Thus, this it follows immediately from \eqref{constant1} that
\begin{equation}\label{position23}
F_{23}(z)F_{32}(z):=-f(z)+(z-1)f'(z).
\end{equation}
From equations \eqref{position13}, \eqref{position12} and \eqref{position23} we have immediately 
\begin{equation}\label{step1}
z\frac{d}{dz}\left(F_{23}F_{32}\right)=z(z-1)\frac{d}{dz}\left(F_{12}F_{21}\right)=-(z-1)\frac{d}{dz}\left( F_{13}F_{31}\right)=z(z-1)f''(z).
\end{equation}
On the other hand, using the equations of the system \eqref{finalODE3} one finds 
$$z\frac{d}{dz}\left(F_{23}F_{32}\right)=z(z-1)\frac{d}{dz}\left(F_{12}F_{21}\right)=-(z-1)\frac{d}{dz}\left( F_{13}F_{31}\right)=F_{21}F_{13}F_{32}+F_{12}F_{31}F_{23}.$$
Combining these equations with \eqref{step1} we obtain 
\begin{equation}\label{painlevevi1}
z(z-1)f''(z)=F_{23}(z)F_{31}(z)F_{12}(z)+F_{13}(z)F_{32}(z)F_{21}(z).
\end{equation}
From \eqref{painlevevi1} we are going to obtain the Painlev\'e VI equation with two parameters exploiting another conservation law. 
Indeed always using equations \eqref{finalODE3} we find that $$\frac{d}{dz}\left(F_{23}(z)F_{31}(z)F_{12}(z)-F_{13}(z)F_{32}(z)F_{21}(z)\right)=0$$
so we set
\begin{equation}\label{constant2}
F_{23}(z)F_{31}(z)F_{12}(z)-F_{13}(z)F_{32}(z)F_{21}(z)=D,
\end{equation}
where $D$ is another constant. 
Squaring \eqref{painlevevi1} we obtain 
$$z^2(z-1)^2 f''^2=\left(F_{13}F_{32}F_{21}\right)^2+\left(F_{12}F_{23}F_{31}\right)^2+2 \left( F_{12}F_{21}F_{13}F_{31}F_{23}F_{32}\right).$$
Squaring \eqref{constant2} and substituting in the previous equation we obtain 
$$z^2(z-1)^2 f''^2=4\left( F_{12}F_{21}F_{13}F_{31}F_{23}F_{32}\right)+D^2.$$
Finally substituting \eqref{position13}, \eqref{position12}, \eqref{position23} in the previous equation, after some straightforward manipulations we obtain 
\begin{equation}\label{painlevevi2}
z^2(z-1)^2\left(f'' \right)^2+4\left[f'(zf'-f)^2-f'^2(zf'-f)\right]+4R^2f'(zf'-f)-4R^2f'^2-D^2=0.
\end{equation}
This proves that given a solution of system \eqref{finalODE3} we can construct a solution of \eqref{painlevevi2}.
 
In order to prove the opposite implication we notice that from \eqref{painlevevi1} and \eqref{constant2} it follows that
\begin{eqnarray}
\label{FFF1}
F_{12}F_{23}F_{31}&=&\f{z(z-1)f''(z)+D}{2},\\
\label{FFF2}
F_{21}F_{13}F_{32}&=&\f{z(z-1)f''(z)-D}{2}.
\end{eqnarray}
Using these identities the system \eqref{finalODE3} can be written as
 \begin{equation}
 \begin{split}
&\frac{d}{dz}\ln{F_{12}}=\frac{1}{z(z-1)}\frac{F_{21}F_{13}F_{32}}{F_{12}F_{21}}=\frac{1}{2z(z-1)}\frac{z(z-1)f''-D}{f'},\cr
&\frac{d}{dz}\ln{F_{13}} =-\frac{1}{z-1}\f{F_{12}F_{23}F_{31}}{F_{13}F_{31}}=-\frac{1}{2(z-1)}\f{z(z-1)f''+D}{f-zf'-R^2},\cr
&\frac{d}{dz}\ln{F_{21}}=\frac{1}{z(z-1)}\f{F_{12}F_{23}F_{31}}{F_{12}F_{21}}=\frac{1}{2z(z-1)}\f{z(z-1)f''+D}{f'},\cr
&\frac{d}{dz}\ln{F_{23}}=\frac{1}{z}\f{F_{21}F_{13}F_{32}}{F_{23}F_{32}}=\frac{1}{2z}\f{z(z-1)f''-D}{(z-1)f'-f},\cr
&\frac{d}{dz}\ln{F_{31}}=-\frac{1}{z-1}\f{F_{21}F_{13}F_{32}}{F_{13}F_{31}}=-\frac{1}{2(z-1)}\f{z(z-1)f''-D}{f-zf'-R^2},\cr
&\frac{d}{dz}\ln{F_{32}}=\frac{1}{z}\f{F_{12}F_{23}F_{31}}{F_{23}F_{32}}=\frac{1}{2z}\f{z(z-1)f''+D}{(z-1)f'-f}.
\end{split}
\end{equation}
Thus we obtain

\begin{eqnarray*}
F_{12}&=&\sqrt{f'}\;\exp\left({- \int_{z_0}^z\left[\frac{1}{2t(t-1)}\frac{D}{f'}\right]\,dt+C_{12}}\right),\\
F_{21}&=&\sqrt{f'}\;\exp\left({\int_{z_0}^z\left[\frac{1}{2t(t-1)}\frac{D}{f'}\right]\,dt+C_{21}}\right),\\
F_{13}&=&\sqrt{f-zf'-R^2}\;\exp\left({-\int_{z_0}^z\left[\frac{1}{2(t-1)}\f{D}{f-tf'-R^2}\right]\,dt+C_{13}}\right),\\
F_{31}&=&\sqrt{f-zf'-R^2}\;\exp\left({\int_{z_0}^z\left[\frac{1}{2(t-1)}\f{D}{f-tf'-R^2}\right]\,dt+C_{31}}\right),\\
F_{23}&=&\sqrt{(z-1)f'-f}\;\exp\left({-\int_{z_0}^z\left[\frac{1}{2t}\f{D}{(t-1)f'-f}\right]\,dt+C_{23}}\right),\\
F_{32}&=&\sqrt{(z-1)f'-f}\;\exp\left({\int_{z_0}^z\left[\frac{1}{2t}\f{D}{(t-1)f'-f}\right]\,dt+C_{32}}\right), 
\end{eqnarray*}
where the constants $C_{ij}$ are integration constants. 
Writing the terms involving square roots as exponentials of integrals and substituting into the original system
 \eqref{finalODE3} we get, after some lengthy computations that the following conditions must hold identically:
\begin{equation}
\begin{split}
&-C_{12}+C_{13}+C_{32}-\alpha+\gamma+[\dots]_1=0\cr
&-C_{21}+C_{23}+C_{31}-\beta+\gamma+[\dots]_2=0\cr
&-C_{13}+C_{12}+C_{23}-\beta+\gamma+[\dots]_2=0\cr
&-C_{31}+C_{32}+C_{21}-\alpha+\gamma+[\dots]_1=0\cr
&-C_{23}+C_{21}+C_{13}-\alpha+\gamma+[\dots]_1=0\cr
&-C_{32}+C_{31}+C_{12}-\beta+\gamma+[\dots]_2=0\\  
\end{split}
\end{equation}
where $\alpha, \beta$ and $\gamma$ are expressed in terms of the initial conditions of the equation \eqref{painlevevi2}  as follows:
\begin{eqnarray*}
\alpha&=&\ln{(f''(z_0)z_0(z_0-1)-D)}\\
\beta&=&\ln{(f''(z_0)z_0(z_0-1)+D)}\\
\gamma&=&\ln{(2\sqrt{f'(z_0)(f(z_0)-z_0f'(z_0)-R^2)(-f(z_0)+(z_0-1)f'(z_0)})},
\end{eqnarray*}
\begin{footnotesize}
\begin{eqnarray*}
[\dots]_1&=&-\int_{z_0}^z\f{d}{dt}\ln{(t(t-1)f''-D)}\,dt+\int_{z_0}^z\f{d}{dt}\ln{[2\sqrt{f'(f-tf'-R^2)((t-1)f'-f)}}]\,dt+\\
&&+D\int_{z_0}^z\f{(f'(t-1)-f)(f-tf'-R^2)+(t-1)f'(f-tf'-R^2)-tf'(f'(t-1)-f)}{2t(t-1)f'(f-tf'-R^2)((t-1)f'-f)}\,dt,\\
\end{eqnarray*}
\end{footnotesize}
and
\begin{footnotesize}
\begin{eqnarray*}
[\dots]_2&=&-\int_{z_0}^z\f{d}{dt}\ln{(t(t-1)f''+D)}\,dt+\int_{z_0}^z\f{d}{dt}\ln{[2\sqrt{f'(f-tf'-R^2)((t-1)f'-f)}}]\,dt+\\
&&-D\int_{z_0}^z\f{(f'(t-1)-f)(f-tf'-R^2)+(t-1)f'(f-tf'-R^2)-tf'(f'(t-1)-f)}{2t(t-1)f'(f-tf'-R^2)((t-1)f'-f)}\,dt.
\end{eqnarray*}
\end{footnotesize}
Using the fact that $f$ is a solution of \eqref{painlevevi2} it is easy to prove that both quantities $[\dots]_1$ and $[\dots]_2$
 vanish. Let us check $[\dots]_1=0$. First of all we observe that
\begin{footnotesize}
\begin{equation}\label{identita}
\begin{split}
f''\left[(f'(t-1)-f)(f-tf'-R^2)+(t-1)f'(f-tf'-R^2)-tf'(f'(t-1)-f)\right]=\\
\f{d}{dt}[f'(f-tf'-R^2)((t-1)f'-f)]
\end{split}
\end{equation}
\end{footnotesize}
Multiplying both numerator and denominator in the third addendum of $[\dots]_1$ by $f''$,
 applying identity \eqref{identita} and taking the common denominator with the second addendum of $[\dots]_1$  we  obtain
\begin{footnotesize}
\begin{eqnarray*}
[\dots]_1&=&-\int_{z_0}^z\f{t(t-1)f'''+(2t-1)f''}{t(t-1)f''-D}\,dt+\\
&&+\int_{z_0}^z\f{(t(t-1)f''+D)\f{d}{dt}[f'(f-tf'-R^2)((t-1)f'-f)]}{2t(t-1)f'(f-tf'-R^2)((t-1)f'-f)f''}\, dt.
\end{eqnarray*}
\end{footnotesize}
Using the equation \eqref{painlevevi2} we obtain the identity
$$f'(f-tf'-R^2)((t-1)f'-f)=\f{1}{4}(t(t-1)f''+D)(t(t-1)f''-D)$$
and as a consequence
\begin{footnotesize}
\begin{eqnarray*}
[\dots]_1&=&-\int_{z_0}^z\f{t(t-1)f'''+(2t-1)f''}{t(t-1)f''-D}\,dt
+\int_{z_0}^z\f{2\f{d}{dt}[f'(f-tf'-R^2)((t-1)f'-f)]}{t(t-1)(t(t-1)f''-D)f''}\,dt+\\
&&=-\int_{z_0}^z\f{2t^2(t-1)^2f''f'''+\f{d}{dt}[t^2(t-1)^2](f'')^2-4\f{d}{dt}[f'(f-tf'-R^2)((t-1)f'-f)]}{2t(t-1)(t(t-1)f''-D)f''}\,dt
\end{eqnarray*}
\end{footnotesize} 
The numerator is nothing but the derivative of the equation \eqref{painlevevi2}. By similar computations we obtain
\begin{footnotesize}
\begin{eqnarray*}
[\dots]_2=-\int_{z_0}^z\f{2t^2(t-1)^2f''f'''+\f{d}{dt}[t^2(t-1)^2](f'')^2
-4\f{d}{dt}[f'(f-tf'-R^2)((t-1)f'-f)]}{2t(t-1)(t(t-1)f''+D)f''}\,dt
\end{eqnarray*}
\end{footnotesize} 
that vanishes for the same reason. It remains to prove that the system
\begin{equation}
\begin{split}
&-C_{12}+C_{13}+C_{32}-\alpha+\gamma=0\cr
&-C_{21}+C_{23}+C_{31}-\beta+\gamma=0\cr
&-C_{13}+C_{12}+C_{23}-\beta+\gamma=0\cr
&-C_{31}+C_{32}+C_{21}-\alpha+\gamma=0\cr
&-C_{23}+C_{21}+C_{13}-\alpha+\gamma=0\cr
&-C_{32}+C_{31}+C_{12}-\beta+\gamma=0\\  
\end{split}
\end{equation}
for the constants $C_{ij}$ admits solutions. It is easy to check that the general solution depends on two arbitrary constants,
 for instance $C_{21}=A$ and $C_{31}=B$:
\begin{equation*}
C_{32}=B-A+\alpha-\gamma,\,C_{13}=-B+\alpha+\beta-2\gamma,\,C_{23}=A-B+\beta-\gamma,\, 
C_{12}=-A+\alpha+\beta-2\gamma. 
\end{equation*}

To conclude let us show that the equation \eqref{painlevevi2} is equivalent to a Painlev\'e VI depending on two parameters.
 For this purpose let us consider the sigma form of the Painlev\'e VI (see \cite{JM81}, Appendix C, Formula C.61): 
\begin{equation}\label{sigmapainleve1}
\frac{d\sigma}{dz}\left(z(z-1)\frac{d^2\sigma}{dz^2}\right)^2+\left(\frac{d\sigma}{dz}\left[2\sigma-(2z-1)\frac{d\sigma}{dz}\right]+v_1v_2v_3v_4 \right)^2=\prod_{k=1}^4 \left(\frac{d\sigma}{dz}+v^2_k \right), 
\end{equation}
where $v_1, v_2, v_3, v_4$ are four parameters suitably related to the other four parameters appearing in the classical form of Painlev\'e VI (see Remark at the end of the proof).

Expanding the products and powers in \eqref{sigmapainleve1} and dividing by $\sigma'$ one gets:
\begin{equation}\label{sigmapainleve2}
\begin{split}
z^2(z-1)^2(\sigma'')^2+4\left[\sigma'(z\sigma'-\sigma)-(\sigma')^2(z\sigma'-\sigma) \right]-4v_1v_2v_3v_4(z\sigma'-\sigma)=\\
(\sigma')^2\left(\sum_{k=1}^4 v_k^2 \right)+\sigma' \left(\sum_{i\neq j}^4v_{i}^2v_{j}^2-2v_1v_2v_3v_4 \right)+\sum_{i\neq j\neq k}^4v_i^2v_j^2v_k^2.
\end{split}
\end{equation} 
Comparing \eqref{painlevevi2} and \eqref{sigmapainleve2}, we see that we need to remove the term $4Rf'(zf'-f),$ since there is no term $\sigma'(z\sigma'-\sigma)$ outside the square bracket. 
Consider the transformation $f=\psi+\frac{R^2}{2}$. Then $f'=\psi'$, $f''=\psi''$ and $zf'-f=z\psi'-\psi-\frac{R^2}{2}$. Substituting in \eqref{painlevevi2} after some straightforward manipulations we obtain 
\begin{equation}\label{sigmapainleve3}
z^2(z-1)^2(\psi'')^2+4\left[ \psi'(z\psi'-\psi)-(\psi')^2(z\psi'-\psi)\right]=2R^2(\psi')^2+R^4\psi'+D^2,
\end{equation}
which can be recognized as a special form of \eqref{sigmapainleve2}, with $\psi=\sigma$ where $v_1v_2v_3v_4=0$. In particular we can choose $v_4=0$ and comparing \eqref{sigmapainleve2} and \eqref{sigmapainleve3} we obtain the following correspondence among parameters:
\begin{equation}\label{relazioni}
\begin{split}
2R^2=v_1^2+v_2^2+v_3^2,\\
R^4=\sum_{i\neq j}^3v_i^2v_j^2,\\
D^2=v_1^2v_2^2v_3^2.
\end{split}
\end{equation} 
Notice that $v_1^2, v_2^2, v_3^2$ are the roots of the cubic polynomial
 $$x^3-\left(v_1^2+v_2^2+v_3^2\right)x^2+\left(\sum_{i\neq j}^3v_i^2v_j^2 \right)x-v_1^2v_2^2v_3^2, $$
 or equivalently, due to the previous relations, roots of the polynomial:
\begin{equation}\label{polynomial}x^3-2R^2x^2+R^4x-D^2.\end{equation}

\endproof

\begin{remark}
The parameters $v_1, v_2, v_3, v_4$ of the complete sigma form of Painlev\'e VI in equation \eqref{sigmapainleve2} are related to the parameters $\alpha, \beta, \gamma, \delta$ of the classical form of Painlev\'e VI:
\begin{equation}\begin{split}\frac{d^2 y}{dz^2}=\frac{1}{2}\left(\frac{1}{y}+\frac{1}{y-1}+\frac{1}{y-z}\right)\left(\frac{dy}{dz} \right)^2-\left( \frac{1}{y}+\frac{1}{y-1}+\frac{1}{y-z}\right)\frac{dy}{dz}+\\
+\frac{y(y-1)(y-z)}{z^2(z-1)^2}\left\{\alpha+\beta \frac{z}{y^2}+\gamma\frac{z-1}{(y-1)^2}+\delta\frac{z(z-1)}{(y-z)^2} \right\}
 \end{split}\end{equation} in the following way: 
$$v_1+v_2=\sqrt{-2\beta}, \quad v_1-v_2=\sqrt{2\gamma}, \quad, v_3+v_4+1=\sqrt{1-2\delta}, \quad v_3-v_4=\sqrt{2\alpha}.$$
However, the relation between a solution of \eqref{sigmapainleve2} and the corresponding solution of classical Painlev\'e equation is more complicated. For explicit formulas, see \cite{O87}.
\end{remark}

\section{An example for arbitrary $n$: the $\epsilon$-system}
Consider the semi-Hamiltonian system (see \cite{ts})
$$u^i_t=\left(u^i-\epsilon\sum_{k=1}^n u^k \right)u^i_x, \quad i=1, \dots, n,$$
known in the literature as $\epsilon$-system. Let us recall that a diagonal system of hydrodynamic type 
$$u^i_t=v^i(u)u^i_x, \quad u=(u^1, \dots, u^n), \quad i=1, \dots n$$
is called semi-Hamiltonian if the characteristic velocities $v^i(u)$ satisfy the following system of equations (here $\partial_j:=\frac{\partial}{\partial u^j}$):
\begin{equation}\label{semihamiltonian}\partial_j \left(\frac{\partial_k v^i}{v^i-v^k }\right)=\partial_k \left( \frac{\partial_j v^i}{v^i-v^j}\right), \quad \forall i\neq j\neq k\neq i.\end{equation}
System \eqref{semihamiltonian} provides the integrability conditions of the following system (among other systems): 
\begin{equation}\label{semih2}\partial_j \ln(\sqrt{g_{ii}})=\frac{\partial_j v^i}{v^j-v^i},\end{equation}
which relates the characteristic velocities of a diagonal system of hydrodynamic type with a class of diagonal metric. 
Let us recall that a metric is said to satisfy the Egorov property if there exists a coordinate system in which the metric is diagonal and potential, namely $g_{ii}=\partial_i \phi$, for a suitable $\phi$. 
Now for $n>2$ the metrics of the form 
\begin{equation}\label{auxiliarymetric}g_{ii}:=\frac{\varphi^i(u^i)}{\left[ \prod_{l\neq i}(u^i-u^l)^2\right]^{\epsilon}},\end{equation}
where $\varphi^i(u^i)$ are arbitrary smooth nowhere vanishing functions of a single variable, do satisfy \eqref{semih2},
 but are not of Egorov type because their rotation coefficients 
\begin{equation}
\label{roteps}
\beta_{ij}=
\left[\f{\prod_{l\ne j}(u^j-u^i)}{\prod_{l\ne i}(u^i-u^l)}\right]^{\epsilon}
\f{\epsilon}{u^i-u^j}\sqrt{\frac{\varphi_j(u^j)}{\varphi_i(u^i)}}
\end{equation}
are not symmetric, and thus the natural connection constructed from these $\beta_{ij}$ does not coincide with the Levi-Civita
 connection associated to any of the metrics of the form \eqref{auxiliarymetric}. 
However, it has been proved in \cite{L2006} that the rotation coefficients \eqref{roteps} 
do satisfy the system (\ref{ED1},\ref{ED2},\ref{ED3}) if the functions $\varphi^i(u^i)$ in \eqref{auxiliarymetric}
 are constants and moreover the Lam\'e coefficients of the metric  (with $\varphi^i(u^i)=1$)
\begin{equation} 
\label{meps}
g_{ii}=\f{1}{[\prod_{l\ne i}(u^i-u^l)]^{2\epsilon}},
\hspace{1 cm}i=1,\dots n,
\end{equation}
satisfies the sytem (\ref{L1},\ref{L2},\ref{L3}) with $d=(n-1)\epsilon$. Indeed
\begin{eqnarray*}
&&\sum_{k=1}^n u^k\d_k g_{ii}=\sum_{k\ne i}\f{2\epsilon u^k}{[(u^i-u^k)\prod_{l\ne i}(u^i-u^l)]^{2\epsilon}}
-\sum_{k\ne i}\f{2\epsilon u^i}{[(u^i-u^k)\prod_{l\ne i}(u^i-u^l)]^{2\epsilon}}=\\
&&\sum_{k\ne i}\f{-2\epsilon}{[\prod_{l\ne i}(u^i-u^l)]^{2\epsilon}}=-2(n-1)\epsilon g_{ii}
\end{eqnarray*}
This means that the connection $\nabla_1$  defined by
\begin{eqnarray*}
\Gamma^{(1)i}_{jk}&=&0\qquad\forall i\ne j\ne k \ne i\\
\Gamma^{(1)i}_{jj}&=&-\Gamma^{(1)i}_{ij}\qquad i\ne j\\
\Gamma^{(1)i}_{ij}&=&\f{H_j}{H_i}\beta_{ij}=\f{\epsilon}{u^i-u^j}\qquad i\ne j\\
\Gamma^{(1)i}_{ii}&=&-\sum_{l\ne i}\Gamma^{(1)i}_{li}
\end{eqnarray*}
and the connection $\nabla_2$ defined by
\begin{eqnarray*}
\Gamma^{(2)i}_{jk}&=&0\qquad\forall i\ne j\ne k \ne i\\
\Gamma^{(2)i}_{jj}&=&-\f{u^i}{u^j}\Gamma^{(2)i}_{ij}\qquad i\ne j\\
\Gamma^{(2)i}_{ij}&=&\f{H_j}{H_i}\beta_{ij}=\f{\epsilon}{u^i-u^j}\qquad i\ne j\\
\Gamma^{(2)i}_{ii}&=&-\sum_{l\ne i}\f{u^l}{u^i}\Gamma^{(2)i}_{li}-\f{1}{u^i},
\end{eqnarray*}
 the product $c^i_{jk}=\delta^i_j\delta^i_k$ and $c^{*i}_{jk}=\frac{1}{u^i}\delta^i_j\delta^i_k$, $e=\sum_{k=1}^n \partial_k$ and $E=\sum_{k=1}^n u^k\partial_k$ define
 a bi-flat semisimple $F$-manifold structure. We thus obtain the following 
 \begin{proposition}
 In any dimension $n$ with $n>2$ the $\epsilon$-system gives rise to a bi-flat semisimple $F$-manifold structure which is not a Frobenius manifold. 
 \end{proposition}  
 
\section{Conclusions}
In this paper we have singled out a class of $F$-manifolds that, although more general than Frobenius manifolds, are sufficiently rich to provide a framework for the study of integrable PDEs of hydrodynamic type, encompassing also those examples like the $\epsilon$-system where the Frobenius manifold theory is not directly applicable. Moreover, due to the requirement that $\nabla_1$ and $\nabla_2$ are hydrodynamically almost equivalent, bi-flat $F$-manifolds are automatically equipped with powerful recursion relations, as we proved in \cite{AL2012}.

Let us briefly comment on the relationships between bi-flat $F$-manifolds and integrable hierarchies of dispersionless PDEs. 
It has been proved in \cite{LPR} that, given a semisimple $F$-manifold with compatible connection, 
 the set of flows of hydrodynamic type
\beq
u^i_t=c^i_{jk}X^ju^k_x,
\eeq
defined by the solutions $X=(X^1,\dots,X^n)$ of the 
equation\beq\label{symmetries}
d_{\nabla}(X\circ)=0
\eeq
commute.

In the case of bi-flat semisimple $F$-manifolds 
$(M,\nabla_1,\nabla_2,\circ,*,e,E)$, using the set-up developed in 
\cite{AL2012}, we illustrate three alternative recurrence schemes to find a countable subset of solutions
 to \eqref{symmetries}.

The starting point is the same for the three procedures. First of all one 
defines the \emph{primary flows} of the hierarchy. These are defined on a semisimple $F$-manifold with compatible flat connection $(M, \circ, \nabla)$. They 
are the flows associated with a frame 
 vector fields $(X_{(1,0)},\dots, X_{(n,0)})$, flat with respect to $\nabla$:
\begin{equation}\label{primflo}
u^i_{t_{(p,0)}}=c^i_{jk}X^k_{(p,0)}u^j_x,
\end{equation}
where $c^i_{jk}$ are the structure constants of $\circ$. 
Starting from the primary flows \eqref{primflo} one can 
introduce  the ``higher flows'' of the hierarchy, defined 
as
\beq
\label{hiflows}
u^i_{t_{(p,\alpha)}}=c^i_{jk}X^j_{(p,\alpha)}u^k_x,
\eeq
by means of one of the following recursive schemes.
\begin{enumerate}
\item The first recurrence procedure can be defined on any semisimple $F$-manifold with compatible connection, namely it holds without a bi-flat $F$-manifold structure:
\beq\label{recrel}
\nabla_{j} X^i_{(p,\alpha)}=c^i_{jk}X^k_{(p,\alpha-1)}.
\eeq
The hierarchy obtained in this way is called \emph{the 
principal hierarchy}. 

\item The second recurrence procedure is available whenever one has a semisimple $F$-manifold $(M, \circ,e)$ with two flat connections, $\nabla_1$ and $\nabla_2$, which are compatible with respect to the {\em same} product $\circ$. Such a structure can always be constructed starting from a semisimple $F$-manifold with compatible connection $(M, \circ, e, \nabla)$, by deforming $\nabla$. Indeed one takes as $\nabla_1$ the undeformed connection and as $\nabla_2$ the connection whose Christoffel symbols are obtained from the Christoffel symbols of $\nabla_1$ adding the structure constants $c^i_{jk}$ of the product $\circ$. 

The recursive procedure is defined via:
\beq\label{rcl}
d_{\nabla^{(1)}}X_{(n+1,\alpha)}=d_{\nabla^{(2)}}X_{(n,\alpha)}.
\eeq
This indeed corresponds to have two 
connections satisfying \eqref{almostcomp} compatible
 with the same product $\circ$\;; connections of this type are called hydrodynamically equivalent connections in the language of \cite{AL2012}. The flows obtained in this 
way are strictly related to the flows of the principal 
hierarchy (see \cite{AL2012}).
\item On the other hand, the third kind of recursive relation requires the presence of a bi-flat semisimple $F$-manifold $(M, \nabla_1, \nabla_2, \circ, *, e, E)$ and is defined via: 
\beq\label{rcl2}
d_{\nabla_{1}}X_{(n+1,\alpha)}=
d_{\nabla_{2}}\left(E\circ X_{(n,\alpha)}\right).
\eeq
Here $\nabla_1$ and $\nabla_2$ are two almost hydrodynamically equivalent flat connections, compatible with $\circ$ and $*$ respectively, where $E$ is the eventual identity relating $\circ$ with $*$. That this is a well-defined recurrence procedure is proved in \cite{AL2012}. 
These recursive relations generalize the standard 
bi-Hamiltonian hierarchy and reduce to it in the case of 
Frobenius manifolds. Let us remark also that one can obtain formally \eqref{rcl}  putting $E:=e$ in \eqref{rcl2}, so that the dual product $*$ coincides with the given product structure $\circ$. 

Notice that the chains of vector
 fields defined in this way might be not independent.
 In this case, following the standard terminology
 for bi-Hamiltonian structures we will say that the
 connections $\nabla_1$ and $\nabla_2$ are 
\emph{resonant}. This happens, for instance, if the two connections 
$\nabla_1$ and $\nabla_2$ have a common flat vector field.
\end{enumerate}

Due to the richness of recursive schemes available on bi-flat $F$-manifolds,
 it is clear that in principle we can construct plenty of new examples of dispersionless integrable hierarchies.
 However, writing down explicitly the equations of these hierarchies in full generality turns out to be a daunting task:
 indeed the computation of the primary flows might be already very difficult. 
\newline

Regarding open problems, we think that there are two main questions of vast scope arising from the comparison between
 bi-flat $F$-manifolds on one side and Frobenius manifolds on the other:
\newline

\emph{The existence of dispersive deformations}.
 In the case of Frobenius manifolds  the principal hierarchy is the dispersionless limit of a full dispersive hierarchy \cite{DZ}. 
 One of the main tools to costruct such a hierarchy
 is the bi-Hamiltonian structure associated to the flat pencil of metrics \cite{du97} defined by
 the invariant metric and the intersection form.
 
 In the more general setting we have considered in
 the present paper the connections $\nabla_1$  and $\nabla_2$ are not related to any known bi-Hamiltonian structure.
 In other words, it is not clear if dispersive deformations preserving integrability do exist and, if they exist, how
 to substitute the powerful bi-Hamiltonian machinery that one has at disposal in the Frobenius case with equally powerful tools. 

Some preliminary results suggest that even in this framework it is possible to introduce a kind of Hamiltonian formalism. We will treat this problem elsewhere. 
\newline
\newline
\emph{The applications of this rich geometric structure to other branches of mathematics}. 
Frobenius manifolds describe moduli space of topological conformal
 field theories and have multiple connections with quantum cohomology, singularity theory,
 Gromov-Witten invariants. Since bi-flat $F$-manifolds have many properties of Frobenius manifolds and they can be constructed from a natural generalization of Darboux-Egorov system it should be natural to expect that they can find applications in similar areas.


\begin{thebibliography}{99}

\bibitem{AVdL}
H. Aratyn, J. van de Leur,
 \emph{Solutions of the Painlev\'e VI equation from reduction of integrable hierarchy in a Grassmannian approach.} Int. Math. Res. Not. IMRN 2008
doi.10.1093/imrn/rnn080

\bibitem{AL2012} 
A. Arsie and P. Lorenzoni \emph{$F$-manifolds with eventual identities, bidifferential calculus and twisted Lenard-Magri chains},
 arXiv:1110.2461.
 
\bibitem{CGM}
R. Conte, A. M. Grundland and M. Musette,
 \emph{A reduction of the resonant three-wave interaction to the generic sixth Painlev\'e equation},
 J. Phys. A {\bf 39} (2006), no. 39, 12115--12127.

\bibitem{Darboux}
G. Darboux, \emph{Le\c{c}ons sur les syst\`emes ortogonaux et les cordonn\'ees curvilignes}, Paris, 1897.

\bibitem{DS}
L.David, I.A.B. Strachan, \emph{Dubrovin's duality for $F$-manifolds with eventual identities},  Advances in Mathematics, 226:5 (2011) 4031--4060.

\bibitem{DS2} 
 L. David,  I.A.B. Strachan, \emph{Symmetries of $F$-manifolds with eventual identities and special families of connections},  arXiv:1103.2045.
 
\bibitem{dn84} B.A. Dubrovin, S.P. Novikov,
 \emph{On Hamiltonian brackets of hydrodynamic type},
 Soviet Math. Dokl. {\bf 279:2} (1984) 294--297.

\bibitem{du93}
B.A. Dubrovin, \emph{Geometry of 2D topological field theories},
in: Integrable Systems and Quantum Groups, Montecatini Terme, 1993.
Editors: M. Francaviglia, S. Greco. Springer Lecture Notes in
Math. {\bf 1620} (1996), pp.\ 120--348.

\bibitem{du97}
B. Dubrovin, \emph{Flat pencils of metrics and Frobenius manifolds}, Integrable systems and algebraic
geometry (Kobe/Kyoto), 1997), 47-72, World Sci. Publishing, River Edge, NJ (1998).

\bibitem{du99}
B. Dubrovin, \emph{Painlev\'e transcendents in two-dimensional topological field theory. The Painlev\'e property},
 287-412, CRM Ser. Math. Phys., Springer, New York, 1999.

\bibitem{DZ} B. Dubrovin, Y. Zhang, \emph{Normal forms of integrable
PDEs, Frobenius manifolds and Gromov-Witten invariants},   math.DG/0108160.

\bibitem{Dad}
B.A. Dubrovin, \emph{On almost duality for Frobenius manifolds}, Geometry, topology
and mathematical physics, 75-132, Amer. Math. Soc. Transl. Ser. 2, {\bf 212}.
Egoroff D.Th., Collected papers on differential geometry, Nauka, Moscow (1970) (in
Russian).

\bibitem{Egorov}
D. Th. Egorov, \emph{Collected papers on differential geometry}, Nauka, Moscow (1970) (in
Russian).

\bibitem{HM}
C. Hertling, Y. Manin, \emph{Weak Frobenius manifolds}, Internat. Math. Res. Notices {\bf 1999}, no. 6, 277--286.

\bibitem{JM81}
M. Jimbo, T. Miwa, \emph{Monodromy perserving deformation of linear ordinary differential equations with rational coefficients. II}, Physica D, Volume 2, Issue 3, June 1981, Pages 407--448.

\bibitem{KK}
S. Kakei, S., T. Kikuchi, \emph{The sixth Painlev\'e equation as similarity reduction of $\mathfrak{gl}$(3) hierarchy},  
Letters in Mathematical Physics {\bf 79} (2007): 221--234. 

\bibitem{K}
A. V. Kitaev, \emph{On similarity reductions of the three-wave resonant system to
the Painlev\'e equations}, J. Phys. A: Math. Gen. 23 (1990), 3543--3553.

\bibitem{LEE}
J. M. Lee, \emph{Manifolds and Differential Geometry}, Graduate Studies in Mathematics, AMS, 2009. 

\bibitem{L2006}
P. Lorenzoni, \emph{Flat bidifferential ideals and semihamiltonian PDEs}, J. Phys. A  {\bf 39} (2006), 
 no. 44, 13701--13715.

\bibitem{LP}
P. Lorenzoni, M. Pedroni, \emph{Natural connections for semi-Hamiltonian systems: 
The case of the $\epsilon$-system}, Letters in Mathematical Physics, {\bf 97} (2011), no. 1, 85--108.

\bibitem{LPR}
P. Lorenzoni, M. Pedroni, A. Raimondo, \emph{$F$-manifolds and integrable systems
 of hydrodynamic type}, Archivum Mathematicum 
{\bf 47} (2011), 163-180.


\bibitem{manin} Y. Manin,
\emph{$F$-manifolds with flat structure and Dubrovin's duality},
Adv. Math. {\bf 198} (2005), no. 1, 5--26.

\bibitem{O87}
K. Okamoto, \emph{Studies on the Painlev\'e Equations I, Sixth Painlev\'e equation $P_{VI}$}, Annali di Matematica Pura e Applicata, 146, 337-381 (1987). 

\bibitem{ps}
M.V. Pavlov. A. Sergyeyev, \emph{Oriented associativity equations and symmetry consistent conjugate curvilinear
 coordinate nets}, arXiv:1204.2514.
 
\bibitem{WW} E. T. Whittaker,  G. N. Watson, \emph{A Course of Modern Analysis}, (1927), Cambridge, UK: Cambridge University Press.
 
 \bibitem{Ricci} G. Ricci \emph{Dei sistemi di congruenze ortogonali in una variet\`a qualunque}, Memorie della Reale Accademia dei Lincei, Classe di Scienze (5), vol. 2 (1896).

\bibitem{ts} S.P. Tsarev,
\emph{The geometry of Hamiltonian systems of hydrodynamic type. The
generalised hodograph transform},
USSR Izv. {\bf 37} (1991) 397--419.

\end{thebibliography}
\end{document}